\documentclass{aa}

\usepackage[english]{babel}

\usepackage{multirow}
\usepackage{amsmath}
\usepackage{amssymb}
\usepackage{breqn}
\usepackage{graphicx}
\usepackage[colorlinks=true, allcolors=blue]{hyperref}
\usepackage{natbib}
\usepackage{rotating}

\usepackage{textcomp}
\usepackage[normalem]{ulem}

\begin{document}

\date{}

\title{Probabilistic classification of X-ray sources applied to \textit{Swift}-XRT and \textit{XMM-Newton} catalogs}



\author{Hugo~Tranin\inst{\ref{ens},\ref{irap}},
Olivier~Godet\inst{\ref{irap}},
Natalie~Webb\inst{\ref{irap}}
\and Daria Primorac\inst{\ref{demcis},\ref{iaps}}
}

\institute{Ecole Normale Sup\'erieure de Paris-Saclay, 61 Av. du Président Wilson, 94235 Cachan Cedex, France\label{ens}
\and
IRAP, Université de Toulouse, CNRS, CNES, 9 avenue du Colonel Roche, 31028 Toulouse, France\\
e-mail: \texttt{hugo.tranin@irap.omp.eu} \label{irap}
\and
Faculty of Electrical Engineering and Computing, University of Zagreb Unska 3, HR-10000 Zagreb, Croatia\label{demcis}
\and
Istituto di Astrofisica e Planetologia Spaziali, IAPS-INAF, Via del Fosso del Cavaliere 100, 00133 Rome, Italy\label{iaps}
}

\abstract
{Serendipitous X-ray surveys have proven to be an efficient way to find rare objects, for example tidal disruption events (TDE), changing-look active galactic nuclei (AGN), binary quasars, ultraluminous X-ray sources (ULXs), and intermediate mass black holes. With the advent of very large X-ray surveys, an automated classification of X-ray sources becomes increasingly valuable.}
{This work proposes a revisited naive Bayes classification of the X-ray sources in the \textit{Swift}-XRT and \textit{XMM-Newton} catalogs into four classes $-$ AGN, stars, X-ray binaries (XRBs), and cataclysmic variables (CVs) $-$ based on their spatial, spectral, and timing properties and their multiwavelength counterparts. An outlier measure is used to identify objects of other natures. The classifier is optimized to maximize the classification performance of a chosen class (here XRBs), and it is adapted to data mining purposes.}
{We augmented the X-ray catalogs with multiwavelength data, source class, and variability properties. We then built a reference sample of about 25000 X-ray sources of known nature. From this sample, the distribution of each property was carefully estimated and taken as reference to assign probabilities of belonging to each class. The classification was then performed on the whole catalog, combining the information from each property.  }
{Using the algorithm on the \textit{Swift} reference sample, we retrieved 99\%, 98\%, 92\%, and 34\% of AGN, stars, XRBs, and CVs, respectively, and the false positive rates are 3\%, 1\%, 9\%, and 15\%. 
Similar results are obtained on \textit{XMM} sources. When applied to a carefully selected test sample, representing 55\% of the X-ray catalog, the classification gives consistent results in terms of distributions of source properties. A substantial fraction of sources not belonging to any class is efficiently retrieved using the outlier measure, as well as AGN and stars with properties deviating from the bulk of their class. Our algorithm is then compared to a random forest method; the two showed similar performances, but the algorithm presented in this paper improved insight into the grounds of each classification.}
{This robust classification method can be tailored to include additional or different source classes and can be applied to other X-ray catalogs. The transparency of the classification compared to other methods makes it a useful tool in the search for homogeneous populations or rare source types, including multi-messenger events. Such a tool will be increasingly valuable with the development of surveys of unprecedented size, such as LSST, SKA, and Athena, and the search for counterparts of multi-messenger events.}
  

\titlerunning{Classification of \textit{Swift \& XMM-Newton} sources}
\authorrunning{Tranin, Godet, Webb \& Primorac }

\maketitle

\section{Introduction}

Changing-look active galactic nuclei (AGN), intermediate-mass black hole (IMBH) candidates, supersoft sources, magnetars, super-Eddington accretors, black holes subject to a tidal disruption event... All of these rare objects were serendipitously found in large X-ray surveys (e.g., \citealt{LaMassa2015}, \citealt{Farrell2009}, \citealt{Henze2012}, \citealt{Zhou2014}, \citealt{Walton2011}, \citealt{Lin2018}), illustrating the wide variety of X-ray emitters. Identifying new, rare sources is essential for extending and deepening our knowledge of the high-energy Universe. These populations can then be used to answer questions regarding the hierarchical evolution of galaxies or how the earliest supermassive black holes formed (e.g., \citealt{Greene2012}). One possible automated way to find such rare objects serendipitously is to search for outliers in the parameter space of the catalog (e.g., \citealt{Lo2014}) within the framework of X-ray source classification. X-ray source classification can also be used for population studies, for example to study the X-ray luminosity function of high-mass X-ray binaries (HMXBs; \citealt{Mineo2012}), to perform quick diagnostics on an object's nature for individual studies, or to spot unstudied objects in unexpected environments (e.g., \citealt{Lin2014}). Another useful application could be to decontaminate samples of ultraluminous or hyperluminous X-ray source candidates (i.e., sources found in the outskirts of nearby galaxies with an X-ray luminosity higher than $10^{39}$ erg~s$^{-1}$), such samples being often contaminated by background AGN and/or foreground stars.

In this context, an automated, efficient way to classify X-ray sources becomes increasingly valuable as X-ray catalogs get larger and larger; for example, the recently launched eROSITA X-ray telescope is predicted to detect about a million sources per year (totaling 3 million AGN by the end of the mission; \citealt{Merloni2012}) and has already detected more than 1 million sources in its first 6 months\footnote{\url{https://www.mpe.mpg.de/7461950/erass1-presskit}}. Historically, X-ray sources were first classified by hand, either by knowing the nature of the visible counterpart (e.g., for a star) or using empirical laws based on the source properties (extent, location in the sky, flux in different bands and colors, or, when available, variability and spectrum; e.g., \citealt{Haberl1999}, \citealt{Prestwich2003}, \citealt{Pineau2011}). Such a manual classification can be very accurate, but it is also tedious, repetitive, and time-consuming. A first way to automate it has been to apply simple threshold rules to source properties (i.e., quantitative selection criteria organized in a decision tree) to separate the parameter space of the sample under study into classes. This approach is often used to distinguish between source natures, typically stars, AGN, and stellar-mass compact objects (e.g., \citealt{Lin2012}). Formally, the choice of the criteria and their application order defines a decision tree. While being a simple, transparent, and physics-based classification model, selection criteria are often quite arbitrary and may be too rough to separate classes properly (classes are often blended for some values of the properties). Besides, this method does not properly handle the presence of missing values among the properties that are used.

Machine learning algorithms have been shown to be efficient in automatically classifying X-ray sources, for example the classification of ROSAT bright sources or \textit{XMM-Newton}  variable sources (\citealt{McGlynn2004}, \citealt{Farrell2015}). For such applications, the random forest algorithm \citep{Breiman2001} is often preferred to other methods, such as neural networks or support vector machines, as it seems to show better results (e.g., \citealt{Arnason2020}) and is theoretically better adapted to X-ray source classification: Indeed, it can be used even with a small number of features (here the source properties) or when they are not normalized, it does not require linearly separable classes, and it is useful even when the training sample is small or unbalanced. In practice, the random forest algorithm consists of the automatic generation of a large number of decision trees similar to the one described above, each trained on a subset of the training sample with a subset of the features (i.e., source properties). Every source is then assigned the most probable identification. Taken individually, each tree performs poorly at classifying sources, but the forest as a whole is able to perform much better.  \cite{Lo2014} showed that \textit{XMM-Newton} variable sources could be reasonably classified thanks to different variability features computed from \textit{XMM-Newton} light curves. The classification efficiency is, however, boosted as soon as the classifier takes other data into account, in particular the multiwavelength information obtained via positional crossmatch between \textit{XMM-Newton} and optical or infrared catalogs (\citealt{Pineau2017}, \citealt{Salvato2018}). While efficient, this classification was applied to a very small fraction ($<1$\%) of the 2XMMi-DR2 catalog \citep{Watson2009}, that is, the variable sources with a good quality flag. Similarly, \cite{Lin2012} considered only the brightest point sources (zero extent and signal-to-noise ratio higher than 20) for their classification, totaling about 2\% of 2XMMi-DR3. We still lack a well-established robust classification method to apply to large fractions of X-ray catalogs in order to achieve the full potential of X-ray classification for data mining. Progress has already been made in this direction, for example by \cite{Pineau2009}, who showed encouraging results of a classification of about 10000 serendipitous sources from 2XMMi-DR2 (which contains about 220000 sources). After applying a principal component analysis (PCA) to reduce the parameter space, they classified these sources by using either a k nearest neighbors (knn) or a kernel density smoothing approach. They also pointed out two limitations: knn provides discrete class probabilities and does not properly handle a bias in the class proportions of the training sample, while kernel density classification requires a large training sample.

Here, inspired from both approaches, we present a probabilistic method for obtaining a refined classification, revisiting the naive Bayes classification algorithm and using the majority of available spatial, timing, and spectral source properties. The combination of all this information, notably obtained by positional crossmatches with multiwavelength catalogs, is likely to produce a robust classification even for faint, non-variable sources. The choice of naive Bayes is motivated by its intuitive nature, as an extension of the rough classification rules used in simplistic decision trees. 
This approach is also transparent as it provides a way to get an insight into the reasons for a particular classification (telling which source properties most influenced it), which is not quite possible for algorithms such as random forest.
We first considered the \textit{Swift}-XRT catalog 2SXPS \citep{Evans2020}, 
taking advantage of the wide sky coverage of the \textit{Swift} Neil Gehrels Observatory \citep{Gehrels2004}, which also enables the variability to be studied on different timescales. After their cross-correlation with optical and infrared catalogs, we applied our algorithm: The classification of each source as an AGN, star, X-ray binary (XRB), or cataclysmic variable (CV) is then based on its location,
X-ray variability, X-ray spectral features (e.g., hardness ratios), and multiwavelength properties. Most X-ray point sources indeed fall into one of these categories, and this choice of classes is further justified in Sect. 2.1.2. Special care is given to the treatment of missing values, which are commonplace in any catalog but can sometimes be used as a piece of information \textit{per se}. We assess the classification performance (retrieval fractions, false positive rates) in various cases, compare it to a random forest algorithm, and finally apply it to the \textit{XMM-Newton} catalog 4XMM-DR10 \citep{Webb2020}.

This paper is organized as follows: In Sect. 2 we describe the X-ray catalogs and what was done to enrich their data; Sect. 3 presents the classification method and how it was assessed; and various miscellaneous results are given in Sect. 4. The comparison with a random forest classification is made in Sect. 5, which also discusses the method and some results, and the main points and outlooks are summarized in Sect. 6.

\section{Use of the X-ray catalogs}

\subsection{\textit{Swift}-XRT catalog: 2SXPS}


The 2SXPS catalog \citep{Evans2020} was used to test our classification method. It consists of 206335 unique sources covering 3790 deg$^2$ of the sky, detected during 1.1 million observations of the \textit{Swift} X-Ray Telescope \citep[XRT;][]{Burrows2004} during its first 13 years of operations (from 2005 January 1 to 2018 August 1). 
The rather uniform sky coverage of 2SXPS and its typical sensitivity of co-added images of $\sim 1.7\times 10^{-13}$ erg~s$^{-1}$~cm$^{-2}$ make it a great database for serendipitous searches. For each source, the location, fluxes in all three bands (0.3$-$1 keV, 1$-$2 keV, and 2$-$10 keV), hardness ratios, quality flags and association with other catalogs are given. Some spectral parameters resulting from an absorbed power law or APEC (Astrophysical Plasma Emission Code, \citealt{Smith2001}, adapted to represent the coronal emission of stars) fit are also present for bright sources, when more than 50 net counts were detected (11\% of 2SXPS). Table 1 summarizes the columns we used in this work and the ones we added. 

In the literature, no classification work has addressed the 2SXPS catalog so far. However, other similar catalogs, such as the \textit{XMM-Newton} serendipitous source catalog (\citealt{Webb2020} in its latest version), are subject to classification works, for example \cite{Lin2012} and \cite{Farrell2015}, respectively, using the "threshold rules" approach and a random forest algorithm, while \cite{Arnason2020} tested several machine learning methods to classify \textit{Chandra} X-ray sources in M31. 
Besides, a study preliminary to this work \citep{Primorac2015} shows that the \textit{Swift} catalog has an interesting potential for X-ray source classification, with a classification based on selection criteria forming a decision tree with three classes: AGN, stars, and stellar-mass compact objects. The last category aggregates objects of differing nature, from XRBs and CVs to isolated white dwarfs, neutron stars and pulsars. Recent progress in survey counts for these objects allows us to split this category into subcategories, for example separating XRBs and CVs, on the basis of their considerably different observational signatures.

In the following, our catalog consists of all 2SXPS sources with a detection flag "Good" or "Reasonable" (ensuring a fraction of spurious sources below $\sim$1\%) and in a clean field (or containing well-modeled stray light), with enough counts to compute a strictly positive flux ($PowFlux$). This sample contains 72\% of the initial catalog (148,438 sources).

For our classification work we enriched the 2SXPS catalog in three steps. First, we completed source properties with new ones including time variability and multiwavelength information, obtained by computation from 2SXPS columns and spatial cross-correlation with catalogs of 
     X-ray, optical and infrared sources (Table 2). Correlations between X-ray catalogs were used to build for each source a light curve of X-ray observations and compute their maximum to minimum flux ratio.
    
Second, we built a sub sample of known sources, henceforth called the "reference sample", by cross-correlating their positions with catalogs of stars, AGN, XRBs, CVs, and ultraluminous X-ray source (ULX) candidates (Table 2). Finally, we selected a test sample of unknown sources, similar to the reference sample in terms of data quality to test our classification under proper conditions.

\begin{table}[h!]
    \centering
    \caption{Columns of the 2SXPS-enriched catalog used in our classification.}
    \resizebox{\columnwidth}{!}{\begin{tabular}{c|c|c}
        Property & Description & Cat.\\\hline
        \multirow{2}{*}{$b$} & \multirow{2}{*}{Galactic latitude} &\multirow{2}{*}{0}\\
        &\\\hline
        \multirow{2}{*}{$HR1$} & Hardness ratio from X-ray&\multirow{2}{*}{1}\\
        &soft-medium bands\\\hline
        \multirow{2}{*}{$HR2$} & Hardness ratio from X-ray&\multirow{2}{*}{1}\\
        &medium-hard bands\\\hline
        \multirow{2}{*}{$FitPowGamma$} & $\Gamma$ of the absorbed &\multirow{2}{*}{1}\\
        & power-law spectral fit \\\hline
        \multirow{2}{*}{$FitAPECkT$} & Temperature of the &\multirow{2}{*}{1}\\
        &APEC spectral fit\\\hline
        \multirow{2}{*}{$IntpPowGamma$} & $\Gamma$ from interpolation &\multirow{2}{*}{1}\\
        & of HR values$^{(1)}$\\\hline
        \multirow{2}{*}{$IntpAPECkT$} & Temperature from &\multirow{2}{*}{1}\\
        &interpolation of HR values\\\hline
        \multirow{2}{*}{$\log_{10}(F_X/F_r)^{(2)}$} & log. of the X-ray to &\multirow{2}{*}{2}\\
        & red band$^{(3)}$ flux ratio \\\hline
        \multirow{2}{*}{$\log_{10}(F_X/F_{W1})^{(2)}$} & log. of the X-ray to &\multirow{2}{*}{2}\\
        & W1-band$^{(4)}$ flux ratio \\\hline
        \multirow{3}{*}{$logFratioSnap^{(2)}$} & log. of the max to mean &\multirow{3}{*}{3}\\
        &flux ratio in most variable\\
        &band from XRT snapshots$^{(5)}$\\\hline
        \multirow{3}{*}{$logFratioObs^{(2)}$} & log. of the max to min &\multirow{3}{*}{3}\\
        &flux ratio in total band\\
        &from X-ray observations$^{(6)}$\\\hline
        \multirow{2}{*}{$FlagSnapBand3^{(2)}$} & Flag set to 0 or 1 if $PvarP$ &\multirow{2}{*}{3}\\
        & $chiSnapshot_{band3}=0$ or 1$^{(7)}$ \\\hline
        \multirow{2}{*}{$L_{X,GLADE~\mathrm{or}~Gaia}^{(2,8,9)}$} & X-ray luminosity from the &\multirow{2}{*}{3}\\
        & distance of the association \\\hline   
        \multirow{2}{*}{$PM_{Gaia}^{(2)}$} & Proper motion of &\multirow{2}{*}{3}\\
        & the Gaia association \\\hline
        \multirow{2}{*}{$distToExtent^{(2,8)}$} & Relative separation to the &\multirow{2}{*}{3}\\
        & center of the associated galaxy\\\hline   
        
    \end{tabular}}
     \tablefoot{The last column gives the category to which the property belongs: 0=location, 
    1=X-ray hardness and luminosity, 
    2=multiwavelength profile, 3=X-ray variability. $^{(1)}$ \cite{Evans2014} simulated a grid of power-law and APEC spectra and computed their corresponding HR values, so they can infer $\Gamma$ and $kT$ values from HR measurements. See \cite{Evans2014} for more details. $^{(2)}$ column added during catalog enrichment. 
    $^{(3)}$ given by the best optical association as explained in the text. The blue band was used likewise to compute a column $\log_{10}(F_X/F_b)$. 
    $^{(4)}$ given by the best infrared counterpart. The W2-band was used likewise to compute 1 extra column. 
    $^{(5)}$ the soft, medium and hard bands were used to compute this column. $^{(6)}$ a positional crossmatch between 2SXPS, 4XMM-DR10 and CSC2 was done to identify sources in common and gather their observations in a single light curve. $^{(7)}$ $PvarPchiSnapshot_{band3}$ is the probability that the source flux is constant between \textit{Swift} snapshots in the hard band. See \cite{Evans2014} for more details. 
    $^{(8)}$ given by the GLADE \citep{Dalya2016} association. $^{(9)}$ given by the catalog of Gaia distances \citep{GaiaDist2021}. 
    }
    \label{tab:1}
\end{table}

\subsubsection{Feature completion}

\begin{table}[]
    \centering
    \caption{Catalogs cross-correlated with 2SXPS and number of matches.}
    \begin{tabular}{c|c|c}
        catalog & Source nature & Matches\\\hline
        Gaia EDR3$^{(2)}$ & Optical sources & 83903\\
        Pan-STARRS DR1$^{(3)}$ & Optical sources & 38228$^{(A)}$\\
        DES DR1$^{(4)}$ & Optical sources & 7534$^{(B)}$\\
        USNO-B1.0$^{(1)}$ & Optical sources & 3812$^{(C)}$\\
        2MASS$^{(7)}$ & Infrared sources &  52989\\
        AllWISE$^{(5)}$ & Infrared sources & 75620$^{(A)}$\\ 
        unWISE$^{(6)}$ & Infrared sources &  15333$^{(B)}$\\
        allWISEagn$^{(8)}$ & AGN & 17033\\
        VV10$^{(9)}$ & AGN & 7017\\
        ASCC$^{(10)}$ & Stars & 4807\\
        Downes 2006$^{(11)}$ & CV &  295\\
        \multirow{2}{*}{Ritter 2014$^{(12)}$} & CV & 339\\
        & LMXB & 45\\
        Liu 2007$^{(13)}$ & LMXB & 40\\
        Kundu 2007$^{(14)}$ & LMXB & 19\\
        Humphrey 2008$^{(15)}$ & LMXB & 9\\
        Zhang 2011$^{(16)}$ & LMXB & 36\\
        Liu 2006$^{(17)}$ & HMXB & 36\\
        Mineo 2012$^{(18)}$ & HMXB & 168\\
        Sazonov 2017$^{(19)}$ & HMXB & 106\\
        BlackCAT $^{(20)}$ & BHB & 18\\
        WATCHDOG $^{(21)}$ & BHB & 27\\
        Liu 2005 $^{(22)}$ & ULX candidates & 312\\
        Walton 2011$^{(23)}$ & ULX candidates & 110\\
        Earnshaw 2019$^{(24)}$ & ULX candidates & 88\\
    \end{tabular}
    \tablefoot{ $^{(A)}$ Does not include the X-ray sources that already have a counterpart identified in the first search (in Gaia for optical sources, 2MASS for infrared sources). $^{(B)}$ Does not include the X-ray sources that already have a counterpart identified in the first and second search (optical: in Gaia or Pan-STARRS, infrared: in 2MASS and AllWISE). $^{(C)}$ Does not include the X-ray sources that already have a counterpart in Gaia, Pan-STARRS, or DES. References: $^{(1)}$ \cite{Page2014} $^{(1)}$ \cite{Monet2003} $^{(2)}$ \cite{Gaia2020} $^{(3)}$ \cite{Chambers2016} $^{(4)}$ \cite{Abbott2018} $^{(5)}$ \cite{Cutri2013} $^{(6)}$ \cite{Schlafly2019} $^{(7)}$ \cite{Cutri2003} $^{(8)}$ \cite{Secrest2015} $^{(9)}$ \cite{Veron2010} $^{(10)}$ \cite{Kharchenko2009} $^{(11)}$ \cite{Downes2006} $^{(12)}$ \cite{Ritter2014} $^{(13)}$ \cite{Liu2007} $^{(14)}$ \cite{Kundu2007} $^{(15)}$ \cite{Humphrey2008} $^{(16)}$ \cite{Zhang2011} $^{(17)}$ \cite{Liu2006} $^{(18)}$ \cite{Mineo2012} $^{(19)}$ \cite{Sazonov2017} $^{(20)}$ \cite{CorralSantana2016} $^{(21)}$ \cite{Tetarenko2016} $^{(22)}$ \cite{Liu2005} $^{(23)}$ \cite{Walton2011} $^{(24)}$ \cite{Earnshaw2019}} 
    \label{tab:2}
\end{table}

While 2SXPS contains cross-correlation information with 14 catalogs, including 2MASS \citep{Cutri2003}, AllWISE \citep{Cutri2013}, and USNO-B1.0 point sources \citep{Monet2003}, its positional matching algorithm uses 
an overestimated X-ray position error ($OrigErr90$; see \cite{Evans2014} for more details) for determining if the two positions agree. 
We therefore performed our own positional crossmatch using TOPCAT version 4.5.1 \citep{Taylor2005}. Besides the catalogs mentioned above, we broadened the search for optical counterparts to the catalogs {Gaia Early Data Release 3} \citep[EDR3;][]{Gaia2020}, Pan-STARRS Data Release 1 \citep[DR1;][]{Chambers2016},  and DES \citep{Abbott2018}, and to the UnWISE catalog \citep{Schlafly2019} for infrared counterparts. However, we kept at most one optical counterpart and one infrared counterpart per X-ray source, thanks to an iterative method: We first searched for optical counterparts in Gaia; then, when X-ray sources have no such counterpart, w{e searched in Pan-STARRS, and eventually in DES and USNO}. This was done to first probe bright magnitudes, before probing deeper in the sky and considering less accurate or not all-sky catalogs. Similarly, in the infrared, we first looked for 2MASS counterparts and then for AllWISE (and finally UnWISE) counterparts when none was found in 2MASS.

{Because of the large sky densities of catalogs of optical and infrared sources, crossmatches between the X-ray catalog and such catalogs were obtained using the Bayesian crossmatching tool NWAY \citep{Salvato2018} in two steps: First we retrieved all counterpart candidates (sources from the second catalog within 15 arcsec of an X-ray source) with the ``CDS Upload X-match'' tool in TOPCAT; then, we ran NWAY to obtain all possible associations and their probability. NWAY allowed the use of optical (r-band) and infrared (W1-band) magnitudes to refine these probabilities. We kept only associations above a certain probability cutoff defined using the dedicated calibration tool in NWAY, to ensure a spurious match rate below 15\% for each catalog. This was done by simulating new X-ray positions (initial positions plus a random offset of up to 1 degree) and performing the crossmatch again. This 15\% spurious match rate is a large improvement compared to the associations provided in 2SXPS, which, for example, have 45\% false positives among AllWISE associations. When an X-ray source had more than one association, only the most likely was kept. NWAY identified about 8\% (4\%) of ambiguous multiple associations in the optical (resp. infrared) catalogs.}

These matches were used as optical and infrared counterparts of 2SXPS sources, in order to compute the ratio of the X-ray flux to their optical or infrared fluxes and use it as another indicator of the source class. Such an indicator was already used, for example, to separate stars and AGN, with simple threshold rules such as ``if $\log(F_X/F_{IR})<-2$ then the source is considered as a star'' (e.g., \citealt{Mikles2006}).\\

The absence of an identified counterpart can 
be taken as information: as expected from their nature, all stars in the reference sample have an optical or infrared counterpart. That is not the case for XRBs ({89}\% -- resp. {76}\% -- have an optical -- resp. infrared -- counterpart), which suggests that a missing value for, for example, $\log(F_X/F_{IR})$ would increase the likelihood of the XRB class compared to the stellar one. The commonly used imputation of such missing values, which are not missing at random, is thus inappropriate: we rather used the fact that they are missing in aid of the classification process.\\

{Another valuable piece of multiwavelength information is the association with the galaxies or nuclear clusters that host the X-ray source. We thus correlated 2SXPS with the GLADE catalog \citep{Dalya2016} of $\sim$2 million galaxies and globular clusters, which is complete to 73 Mpc and relatively complete (>50\%) even at 300 Mpc. Because it is designed to search for counterparts of gravitational wave events, GLADE contains the distance of all its objects, enabling the computation of the source luminosity. We selected X-ray sources as soon as their X-ray error circle overlapped with the ellipse representing a galaxy extent (parameterized in GLADE by the galaxy center, a major axis -- which is $D_{25}$, the diameter of the $Bmag = 25.1\, \mathrm{mag/arcsec}^2$ isophote level -- a minor axis and a position angle), using the ``Sky Ellipses'' algorithm in TOPCAT.}

We added extra columns to estimate the source variability. This could be done directly from the light curves, as in \cite{Lo2014}, but as a first approach we computed only the logarithm of the ratio between maximum and mean X-ray fluxes, in the most variable energy bands (among soft 0.3--1~keV, medium 1--2~keV, and hard 2--10~keV bands). This ratio was estimated by the quantity $PeakRate_{band i}/Rate_{band i}$ or $PeakRate_{band i}/UL_{band i}$ if $Rate_{band i}$ is zero, where $UL_{band_i}$ stands for the 3$\sigma$ upper limit on the count rate, given in 2SXPS. Additionally, we combined the catalogs of detections from \textit{Swift}, \textit{XMM-Newton} and \textit{Chandra} to compute the flux ratio between all observations of a single source. A last piece of information about variability is also present in the 2SXPS catalog: For example $PvarPchiSnapshot\_band3$ is the probability that the source count rate in the hard band  (2--10 keV) does not vary between snapshots. However, in practice this quantity often takes the value 0 or 1, and if not, it seems difficult to exploit as a probability. We preferred to use a flag $FlagSnap\_band3$ to tell if this probability equals 0 (flag set as 0), 1 (flag set as 1), or an intermediate value (flag set as missing value).

\subsubsection{The reference sample}

As any catalog of X-ray sources, we expect a wide diversity of source natures in 2SXPS, with typically AGN and stars as the most dominant populations. To implement a classification, a sample of reference objects for which this nature is already known is required. Since no such information is provided in 2SXPS, we cross-correlated it with catalogs of identified AGN, stars and other X-ray source types, as detailed in Table 2, in order to build our reference sample. It should be noted that the All-Sky Compiled Catalog (ASCC) and the allWISEagn catalog may be contaminated by CVs and stars, respectively, for about 0.4\% of their sources, as inferred from matches with Simbad sources. The cross-correlation was done through the positional crossmatch algorithm implemented in TOPCAT, using the \textit{Err90} (90\% confidence) error for 2SXPS and the $1\sigma$ error given in the secondary catalog, if given. When this error was missing, it was replaced with 0.1 arcsec for allWISEagn, the typical position error of WISE sources, and 1 arcsec for other catalogs, for the case of VV10 and sparser, less studied catalogs of low-mass x-ray binaries (LMXBs), HMXBs, and CVs. For AGN, we combined the catalogs of \citealt{Veron2010} (VV10) 
and \citealt{Secrest2015} (allWISEagn), leading to the identification of 20819 AGN in 2SXPS. For stars, 5232 matches were obtained between 2SXPS and the ASCC (\citealt{Kharchenko2009}, combining several large catalogs of stars with proper motion information). Another important type of X-ray source is the stellar-mass compact objects, especially those in an accretion state. As a compromise between their diversity of signatures and the number counts of each class (which has to be high enough for statistics purpose), we chose to keep only two classes for stellar-mass compact objects, one gathering LMXBs and HMXBs into the class ``XRB'' and the other made up of all kinds of CVs. 
The reference sample is made of all these identified sources, after removing the spurious double identifications that happen, for example, in areas of high source crowding when \textit{Err90} is large. They were identified when present in several catalogs corresponding to different natures. 26 potential AGN, 51 potential stars, 29 potential XRBs, and 32 potential CVs were discarded in this process (totaling 69 sources). 
Last but not least, we kept only "good quality" sources in the training sample, in the sense that they could be classified manually (see Sect. 2.1.3 for details), in order to build the classifier on classifiable sources. The resulting sample contains 
25160 sources including 78\% of AGN 
(19708), 19\% of stars 
4737), 1.4\% of XRBs 
(356), and 1.4\% of CVs 
(359). Figure 1 shows their galactic latitude and X-ray to optical (r-band) flux ratio distributions per source class. These properties can individually help in manual source recognition up to a certain point, but this is limited by overlaps on rather large intervals between the distributions of different classes.

\begin{figure}
    \centering
    \includegraphics[width=8cm]{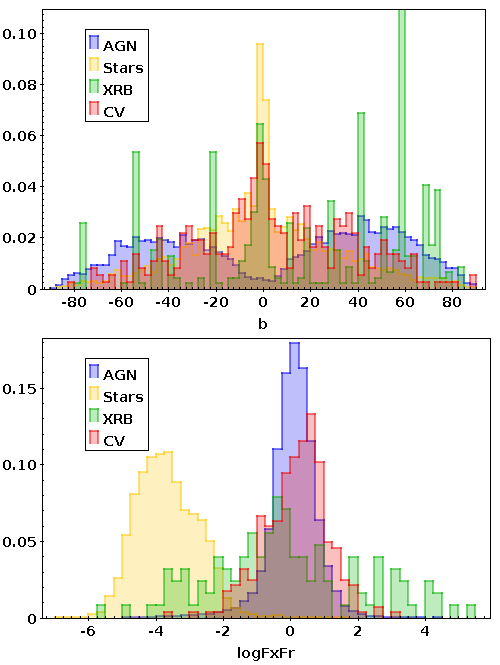}
    \caption{Normalized histograms of the galactic latitude $b$ (top) and the X-ray to r-band flux ratio $\log_{10}(F_X/F_r)$ (bottom) for the sources of the reference sample. The four classes, AGN, star, XRB, and CV, are shown. The green spikes present in the first panel correspond to latitudes of galaxies in which many XRBs were found in the literature (e.g., M31, NGC5457, and NGC1399).}
    \label{fig:1}
\end{figure}

\subsubsection{The test sample}


As every classification based on a training sample, our model must be applied to a test sample with similar properties if we want its results to be reliable. For example, the reference sample has naturally a flux distribution shifted to the bright end compared with the bulk of \textit{Swift} sources, because faint sources are often too faint to be identified. While part of our goal is to provide a classifier that can give insight into the nature  of the majority of the sources, including the faintest ones, it appears unlikely that the whole catalog can be classified currently. We thus chose to select the test sample by looking at a certain number of rules (and excluding sources in the reference sample): (1) the X-ray source must have a counterpart in an infrared catalog; (2) the X-ray source must have a counterpart in an optical catalog; (3) the X-ray source must have been detected several times, regardless of the X-ray telescope; and (4) the X-ray source must have a signal-to-noise ratio higher than 10 or its spectrum must have been acquired. 

Rather than requiring each source of the test sample to follow all four rules, we select them as soon as they follow at least two of these rules. The number of rules they follow is then stored for later use, to consider several test samples of different quality. However, we do not necessarily expect a better-quality classification for the rare sources following all four rules, because they are similar to very few sources of the training sample on which the classification is based. Consequently our whole test sample contains 115361 unique sources (56\% of 2SXPS) and the best-quality sample contains 4\% of 2SXPS.

The requirement of each source to follow two rules is motivated by a case-by-case analysis: a manual classification is relevant, and our algorithm seems to work well as soon as several categories (i.e., variability, hardness, and multiwavelength data) have an information. We suggest that this approach can be considered as necessary and sufficient to build up a test sample, since sources that could be reliably classified and not following at least two rules are rare. This happens, for example, for the few sources that have neither a counterpart nor multiple detections but do have a good spectrum.

 When the same selection principle is applied to the reference sample, only 1150 sources (<5\%) are dismissed.

\subsection{\textit{XMM-Newton} catalog: 4XMM-DR10}

We also considered one of the latest release of the \textit{XMM-Newton} serendipitous source catalog, 4XMM-DR10 \citep{Webb2020}. {In order to focus on reliable and point-like sources, our sample consisted of sources that have no detected extent and are considered to be clean (summary flag < 3) in at least one detection.} Compared to \textit{Swift}-XRT, the \textit{XMM-Newton} telescope has a better sensitivity and a better spectral resolution but a smaller sky coverage, a smaller proportion of sources having been followed up on. There is also less variability information in the \textit{XMM} catalog than in the \textit{Swift} one, for example the peak flux inside any observation or in each energy band. The same procedure as the previous section was applied to 4XMM-DR10. 
By contrast, hardness information is characterized by four hardness ratios instead of two due to the five energy bands used in the catalog. The technique used to find optical and infrared counterparts was similar. 
 Figure A.1 in the appendix shows the distributions of different properties for each class, just as Fig. 2 but for 4XMM-DR10. 

While the reference sample we build is of size similar to 2SXPS (25000 objects), the proportion of each type of X-ray sources in 4XMM-DR10 is slightly different: AGN, stars, XRBs, and CVs represent 72\%, 25\%, 2.1\%, and 1.0\% of the reference sample, respectively (see the fifth line of Table 5 for the detailed number counts). 

Regarding the test sample, in 4XMM-DR10, 315378 sources (55\% of the whole catalog) follow at least two rules and 47544 sources (8\%) follow the four rules. In proportion to the whole catalog, this is very similar to 2SXPS, though due to different reasons: 52\% of 2SXPS have multiple X-ray detections and only 12\% have a spectrum or a large signal-to-noise ratio. These percentages become 28\% and 33\% in the case of 4XMM-DR10. This is notably because 
of the frequent follow-up of \textit{Swift} sources, while \textit{XMM} has a better sensitivity.

\section{Classification method}

In the following we consider a supervised classification method, that is, we assume the existence of a sample of known sources (called the reference sample, or training sample), thanks to which the classification method can be developed and tested.

Our probabilistic classification method can be summarized as follows. First, we estimate a probability density of each class for every property, from the normalized histogram of the same property for identified objects of the reference sample.

Second, these probability densities are used to compute the likelihoods that the source to classify belongs to a certain class \textit{given} each of its properties. When a property value is missing, and if the property is not missing at random, this likelihood is replaced by the probability that a source of this class has this value missing. It is computed as the frequency of such sources among all known sources of this class.
    
    Third, to compute the probability that the source belongs to the class given its properties, we use the Bayes rule. The likelihood term is here computed as the weighted product of likelihoods given by the previous step: the weights are optimized in such a way to maximize the classification performance of one particular class (XRB in our case).

Fourth, the class giving the higher probability is assigned to this source, and the probabilities of all classes are recorded. Fifth, the classification performance is assessed by cross-validation on the reference sample.

The following sections present an intuitive reasoning leading to this classification, the kernel density estimation (KDE) it needs as input, how it is fine-tuned to maximize its performance for a chosen class, and how it can be evaluated. 

\subsection{Theory}

When dealing with an unknown source we want to classify, the manual way is to compare its properties with prior knowledge (i.e. here the distributions of the reference sample). For instance, we can consider a source with a value of the galactic latitude $b=50^{\circ}$ and $\log_{10}(F_X/F_r)=-3$ (logarithm of the X-ray to optical flux ratio). Assuming we want to classify it either as an AGN or as a star, the $b$-value tends to rule out the stellar nature while the $F_X/F_r$-value tends to support it (Fig. 1). In the absence of other information, the choice is then either arbitrary or based on probabilities: The class to prefer is the one that gives the highest density in the $b-\log_{10}(F_X/F_r)$ plane at these coordinates, or the highest product $d_b(50^{\circ})\times d_{\log_{10}(F_X/F_r)}(-3)$ if the two properties are considered independent of each other. The density $d_b$ (respectively $d_{\log_{10}(F_X/F_r)}$) is hence considered as the likelihood for the source to belong to a certain class given its $b$ ($\log_{10}(F_X/F_r)$) value. If we assume then that there are 60\% of AGN and 40\% of stars in our catalog, following the Bayes rule with priors $\mathcal{P}(\mathrm{AGN})$ and $\mathcal{P}(\mathrm{Star})$, the probability that this source is an AGN given its data $D$ follows:

\begin{align*}
&\mathbb{P}(\mathrm{AGN}|D)= \frac{\mathcal{P}(\mathrm{AGN}) \mathcal{L}(\mathrm{AGN}|D)}{\mathcal{P}(\mathrm{AGN}) \mathcal{L}(\mathrm{AGN}|D) + \mathcal{P}(Star) \mathcal{L}(Star|D)}\\\\
& = \frac{0.6~d_{b}^{AGN}(50^{\circ})d_{F_X/F_r}^{AGN}(-3)}{0.6~ d_{b}^{AGN}(50^{\circ})d_{F_X/F_r}^{AGN}(-3) + 0.4~d_{b}^{Star}(50^{\circ})d_{F_X/F_r}^{Star}(-3)}\\\\
&\approx 15\%, 
\end{align*}

\noindent where $d^{AGN}$ and $d^{Star}$ are the KDE of the AGN and star distributions shown in Fig. 1. This is basically the concept of a naive Bayes classifier \citep{Murphy2006}. This model was chosen in order to address the issues of classifications based on a decision tree dividing the parameter space with rough classification rules (e.g., \citealt{Lin2012}), but also to still keep a simple and transparent method. In practice we considered all relevant properties of a source at once and we assigned it the probability to be a star, an AGN, an XRB, or a CV (i.e., the four classes we consider in this work). We used an outlier measure (detailed in Sect. 3.2) to be able to spot sources not belonging to any of these classes. A naive Bayes classifier relies upon the assumption of independent features, which is clearly false if we consider all columns cited in Table 1 at once. Any couple of highly correlated features means virtually the double-counting of a feature by the classifier. Therefore, we separated catalog columns into categories we assumed to be more or less independent: location, hardness ratios, multiwavelength profile and variability, labeled from 1 to 5 in the last column of Table 1. Even if these categories are actually not independent, this is corrected by the classification optimization, done by weighting each category (as detailed below). 
Thus, the likelihood of the data given a source class $\mathrm{c}$, $\mathcal{L}(data|\mathrm{c})$, becomes 

\begin{align*}
\mathcal{L}(data|\mathrm{c})&=\mathcal{L}(location|\mathrm{c})\times\mathcal{L}(HR|\mathrm{c})\\
&\times~\mathcal{L}(mutliwavelength|\mathrm{c})\times\mathcal{L}(variability|\mathrm{c})~~(1)
\end{align*}

\noindent with, for example,
\begin{align*}
\mathcal{L}(HR|\mathrm{c})=&d_{HR1}^{~\mathrm{c}}(HR1) d_{HR2}^{~\mathrm{c}}(HR2)\\
&\times~\mathcal{L}(HR1\text{ present}| \mathrm{c})
\mathcal{L}(HR2\text{ present}| \mathrm{c}),~~~~~~(2) 
\end{align*}

\noindent where HR1 (respectively HR2) is the hardness ratio between the soft and medium (medium and hard) energy bands. $\mathcal{L}(HR1\text{ present}| \mathrm{c})$ is the probability that the value is present given that the source belongs to class c. It is computed as the frequency of such sources in class c of the reference sample. If the source misses HR1 and HR2 values, then $\mathcal{L}(HR|\mathrm{c})$ becomes

$$\mathcal{L}(HR|\mathrm{c})=\Big[\mathcal{L}(HR1\text{ missing}| \mathrm{c})\mathcal{L}(HR2\text{ missing}| \mathrm{c})\Big]^{1/2}.$$

When the property $p$ under consideration is missing at random (which is the case of HR1 and HR2 in practice), $\mathcal{L}(p\text{ present}| \mathrm{c})$ and $\mathcal{L}(p\text{ missing}| \mathrm{c})$ do not depend on class c, so this term can be safely replaced by a constant (it is canceled out when computing posterior probabilities).

For data mining purpose, in this context of very unbalanced classes, it is tempting to have a flexible classification, with maximized performance for the class of highest interest. We chose to optimize the classification for XRBs, for the sake of decontaminating extragalactic X-ray sources. Indeed in the hunt for, for example, hyperluminous X-ray source (HLX) candidates (which was the initial motivation of this work), it is valuable to have both a high retrieval fraction and a low false positive rate for XRBs as opposed to their possible contaminants. A fine-tuning of the classification was done by weighing the likelihood of each category by an ad hoc coefficient $\alpha$ as an exponent. Eventually, the probability for the source to belong to a certain class c was computed as

\begin{align*}
&\mathbb{P}(\mathrm{c}|data)=\\
&\frac{\mathcal{P}(\mathrm{c})\times\left(\prod_{t\in\{\text{cat}\}}\mathcal{L}(t|\mathrm{c})^{\alpha_t}\right)^{1/\sum_{t\in\{\text{cat}\}}\alpha_t}}{\sum_{C\in\{\text{classes}\}}\mathcal{P}(C)\times\left(\prod_{t\in\{\text{cat}\}} \mathcal{L}(t|C)^{\alpha_t}\right)^{1/\sum_{t\in\{\text{cat}\}}\alpha_t}},~~(3)
\end{align*}\\

\noindent where the priors $\mathcal{P}$ of AGN, stars, XRBs, and CVs  are set to 66\%, 25\%, 7\%, and 2\%, respectively. These values are decisive for classification results, and this choice, based on a rough approximation of the proportions we can expect in a standard X-ray catalog, will be discussed in Sect. 5.4. In practice, priors put more stringent constraints for an object to be classified as a rare type: it is thus a way to avoid high false positive rates, at the expense of the $recall$ rates.\\


\subsection{Outlier measure}

Some objects do not belong to any of the classes mentioned above. Such objects can have anomalous X-ray and multiwavelength properties, in the sense that they lie outside the bulk of the distributions of AGN, stars, XRBs, and CVs. Alternatively they can have no class consistent with all their properties at once. We implemented an outlier measure to identify some of them: It consists of minus the logarithm of the product of likelihoods for the predicted class c (Eq. 4). The lower the property distribution of the output class c at a given property value, the lower $\mathcal{L}(t|\mathrm{c})$ (where $t$ is the category of the considered property) and the higher the outlier measure:

$$~O.M.= -\log\left(\mathcal{P}(\mathrm{c})\times\prod_{t\in\{\text{cat}\}}\mathcal{L}(t|\mathrm{c})^{\alpha_t/\sum_{t\in\{\text{cat}\}}\alpha_t}\right).~~~(4)$$

\subsection{Kernel density estimation of the property distributions}

As described in Sect. 3.1, we need to compute the densities underlying the distributions of each property for each class. A common estimate is to use the histogram of the distribution directly, but this presents a number of drawbacks: The shape of the histogram is highly sensitive to the chosen bin width, leading to spurious features if the bin width is too small compared to the size of the data set or a loss of features if it is too large, but also to the chosen bin phase (i.e., the zero point for binning). Methods exist to infer the optimal histogram bin size (e.g., \citealt{Izenman1991}); however, they depend on the size of the sample, meaning a different bin size for each class, which can also bias the classification. A much more robust and modern technique to estimate the density functions underlying our sample distributions is the KDE (\citealt{Sheather2004}).  This nonparametric approach offers a great flexibility in effectively modeling probability density functions from a sampled data set, and thus has become increasingly popular \citep{Botev2010}. Basically, KDE consists in computing the estimate of the density $f$ as the sum of contributions of each individual data point, where the ``contribution'' of a point is a bell-like curve (the kernel) centered on its value:

$$\widehat{f}(x)=\sum_{observations} K(\frac{x-observation}{h}).$$

The kernel was chosen to be exponential, $K(x;h)\propto \exp(-|x|/h)$, as a compromise to detect narrow features while keeping a smooth global density shape. Here, $h$ represents the bandwidth of the kernel: the higher the bandwidth, the smoother the estimate of $f$, with potential loss of features. The Silverman \citeyearpar{Silverman1986} rule of thumb, and a manual inspection of our sample distributions and their KDEs, led to us choosing a bandwidth such that $bandwidth=0.2\Delta x \cdot n^{-1/5}$, where $\Delta x$ is the range of the values of the property $x$ and $n$ is the number of sources that have a non-missing $x$ value in the reference sample.

To solve the problem of zero probabilities, arising when a source to classify has a property for which the KDE is null, we applied Lidstone smoothing \citep{Raschka2017} by adding a constant offset to the corresponding density estimates: for a given property and a given class, this offset was 0.01 times the number of sources belonging to this class and not having this property missing. Another issue was the sparsity of certain property distributions, due to the physical quantity they represent. For instance, the ratio of the X-ray to optical fluxes had to be considered in logarithmic scale to correctly sample the range which can be decades. Other such examples are the spectral fitting parameters; for example, $FitPowGamma$ and $FitAPECkT$ in 2SXPS have most of their data points in a low-value range ($-1$ to $10$ and $0$ to $20$ keV, respectively), but some of them have significantly larger values  \citep[up to 100 and 64 keV, respectively, due to the spectral fitting algorithm not converging to a proper solution;][]{Evans2014}. These fitted spectral properties were thus considered {in logarithmic scale} to squeeze their values into a smaller interval.

Figure 2 shows four examples of KDEs applied to our 2SXPS data set, on galactic latitude $b$, $\log_{10}(F_X/F_r)$, $HR1,$ and $logFratioSnap$ (Table 1). We see that classes are somewhat efficiently separated by the combination of these four properties. We inspected each output density estimation to avoid any bias: Only the galactic latitude showed a clear bias for the XRB class, with high isolated peaks corresponding to the latitude of nearby galaxies in which many XRBs were identified. This density was corrected by replacing half of it by a uniform distribution (i.e., proportional to $cos(b)$), to prevent the classification from giving less credibility to the XRB class when handling extragalactic sources outside these few galaxies. A closer look at the other density estimations we obtained is provided in Appendix A. 

\begin{figure}
    \centering
    \includegraphics[width=8cm]{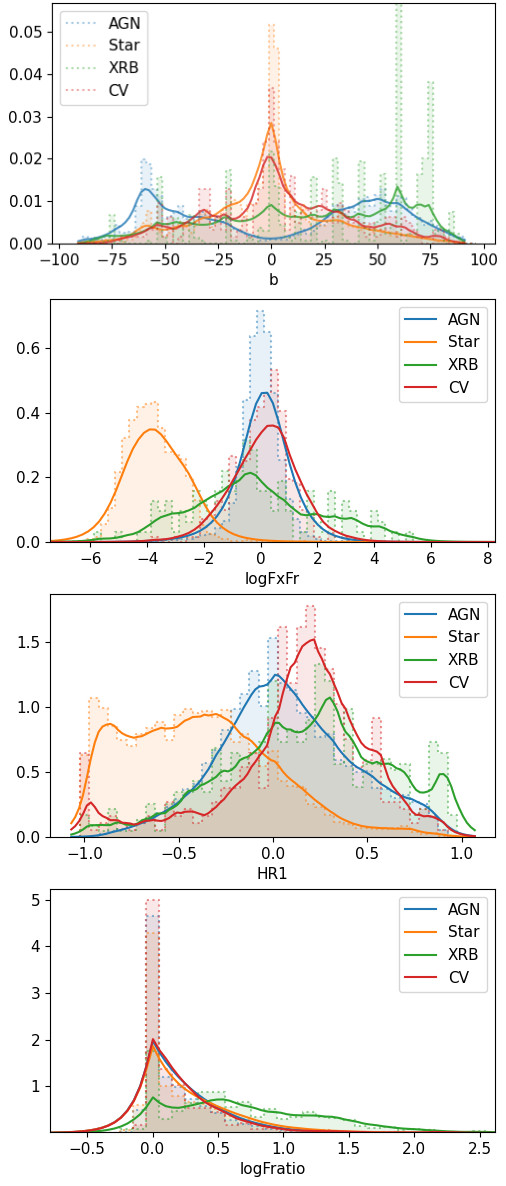}
    \caption{Comparisons between the distributions of different properties in the reference sample of 2SXPS and their KDE. From top to bottom: galactic latitude ($b$), X-ray to r-band flux ratio ($\log_{10}(F_X/F_r)$), hardness ratio between soft and medium X-ray bands ($HR1$), and logarithm of the max over mean flux ratio in the most variable energy band ($logFratioSnap$). The y axis is in arbitrary units.}
    \label{fig:2}
\end{figure}

\subsection{Fine-tuning of the model}

We optimized our classification to maximize its performance regarding the XRB class: coefficients 
$\alpha_t$ of Eq. (3) are tuned to maximize the $f_1$ score of this class, namely,  $f_1=2(recall^{-1}+precision^{-1})^{-1}$, where $recall$ is the fraction of actual XRBs retrieved as XRBs by the classification algorithm, and $precision$ is the fraction of true positives among the sources classified as XRBs. To properly estimate $precision$, the classification was performed on a reference sample with realistic proportions (i.e., the ones we used as priors in Eq. (3)). 
Such an optimization is theoretically motivated by the fact that some categories of properties have more discriminant power than other. Moreover, if two categories are strongly correlated (for example if we consider hardness ratios and spectral fitting properties in two separate categories), the Bayes classifier assumption of independence will lead to the double count of such properties (they will be considered as overly important): setting coefficients $\alpha_t$ to an optimal value is a solution to this issue.

For this optimization, random sets of coefficients were uniformly generated in the range 
$0-10$ for each $\alpha_t$, the classification was run on a properly proportioned reference sample (i.e., a subset respecting the proportion of each class) and the $f_1$ score was computed. We made use of a differential evolution algorithm \citep{Storn1997} to derive the best set of coefficients: this optimization method available in Scipy 1.4.1 has the advantage that it does not require the problem to be differentiable and to probe a very large space of coefficient solutions. Besides, it is barely limited by the potential presence of local minima. We saved the coefficient set at each iteration of the algorithm for manual validation, and to gain insight into the $precision$--$recall$ compromise arising when fine-tuning our classification model.

\begin{figure}
    \centering
    \includegraphics[width=8cm]{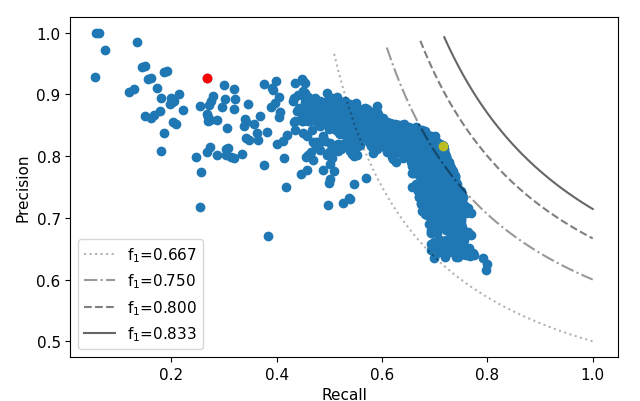}
    \caption{$precision$ versus $recall$ of the classification run at each iteration of the differential evolution algorithm. Contours of the $f_1$ score are shown for reference. The algorithm converges to the point closer to the contour $f_1=0.8$. The red and green dots correspond to the first and last iterations, respectively.}
    \label{fig:3}
\end{figure}

{The final coefficients after convergence of the algorithm are shown in Table 3. Interestingly, the most decisive properties are the ones from the ``location'' category: this is physically meaningful since our classification is optimized for XRBs, which overlap many of the AGN properties, but can be roughly separated using their location inside the hosting galaxy and their luminosity.}

\begin{table}[]
    \centering
     \caption{{Parameters of the classifier after optimization.}}
    \begin{tabular}{c|c}
        \multicolumn{2}{c}{2SXPS}\\\hline
        Parameter & Value \\\hline
        Priors $\mathcal{P}$ (AGN, star, XRB, CV) & 0.66, 0.25, 0.07, 0.02 \\
        Coefficient $\alpha_{\mathrm{location}}$ & 8.8\\
        Coefficient $\alpha_{\mathrm{hardness}}$ & 7.3\\
        Coefficient $\alpha_{\mathrm{multiwavelength}}$ & 2.1\\
        Coefficient $\alpha_{\mathrm{variability}}$ & 3.9\\\\\hline
        \multicolumn{2}{c}{4XMM-DR10}\\\hline
        Parameter & Value \\\hline
        Priors $\mathcal{P}$ (AGN, star, XRB, CV) & 0.66, 0.25, 0.07, 0.02 \\
        Coefficient $\alpha_{\mathrm{location}}$ & 7.5\\
        Coefficient $\alpha_{\mathrm{hardness}}$ & 3.2\\
        Coefficient $\alpha_{\mathrm{multiwavelength}}$ & 2.0\\
        Coefficient $\alpha_{\mathrm{variability}}$ & 6.0\\\hline
    \end{tabular}
    \label{tab:3}
\end{table}

Another result shown by Fig. 3 is that the choice of weighting coefficients is always a trade-off between $recall$ and $precision$: Some sets of coefficients can have fewer false positives among sources classified as XRBs (e.g., only 10\%), but they will be unable to correctly retrieve the majority of XRBs. Conversely, some known XRBs seem to be never retrieved by any choice of coefficients: they were manually inspected to try to understand the reasons for that limitation. This is detailed in Sect. 4.1, as well as the study of XRB false positives.

\subsection{Classification performance assessment}

To evaluate our classification method, we used two approaches: First, we looked into common classification metrics obtained from the reference sample. Then, we analyzed the results obtained on the test sample, in terms of proportions and property distribution of each predicted class.

The most direct way to visualize classification results is by looking at the table showing the number counts of each output class for each input class, called the confusion matrix. A classifier is also characterized by its ability to retrieve most objects from each input class ($recall$) and to limit the number of false positives in each output class ($precision$). The $f_1$ score is often used to combine these two measures (e.g., \citealt{Lukic2018}), and the average of $f_1$ scores over all classes summarizes the classifier performance in a single number. While these metrics may be difficult to interpret, they still can be compared to baselines such as their values for a ``random guess'' classifier or the classifier choosing always the largest class. Usually, the reference sample is split into a training sample on which the classifier is built (i.e., the KDEs here) and a validation sample on which it is assessed. This is useful for methods subject to important overfitting of the reference sample (e.g., neural networks). We preferred to use all the reference sample for training, since nonrepresentative features are smoothed during the KDE process. The classification was also assessed on the reference sample, computing $precision$ values with a formula calibrated to reflect a sample that has the same class proportions as the priors (which is supposed to be the case in the test sample), namely$$precision(AGN)=\frac{N_{AGN\rightarrow AGN}\times\mathcal{P}(AGN)/f_{AGN}}{\sum_{C\in\{\text{classes}\}} N_{C\rightarrow AGN}\times\mathcal{P}(C)/f_{C}},~~(5)$$
where $f_{C}$ is the frequency of class $C$ in the reference sample, and $N_{C\rightarrow AGN}$ is the number of sources of class $C$ that were classified as AGN.

We used all these metrics to assess our classification performance. In practice, for the elimination of, for example, HLX contaminants in a sample of candidates, we want the foreground stars and background AGN to be correctly classified: this is limited by the false positive rate ($1 - precision$) obtained for XRBs. 
However, we also want the actual HLXs not to be classified as contaminants: That is the point of having a high retrieval fraction. Sect. 4.1 details the results we obtain for different choices of classification fine-tuning.

Regardless of the classification performance on the reference sample, it is crucial to study its results on the test sample of unknown sources, in order to characterize its biases. This is possible, for example by looking at some individual sources and comparing the classifier result to the one inferred from a manual analysis, or by comparing the statistics of output classes with the ones of the reference sample. For instance, 
%
the property distributions for each output class should be consistent with the KDEs. The results of these tests on the classification of 2SXPS test samples are summarized in Sect. 4.2.

\section{Results}

In the following we analyze the results of our classification applied to either the reference sample (Sect. 4.1) or the test samples of 2SXPS (Sects. 4.2, 4.4, and 4.5) and 4XMM-DR10 (Sects. 4.2 and 4.3).

\subsection{Results on the 2SXPS reference sample}
When applied to the 2SXPS reference sample, composed of {19708} AGN, {4737} stars, {356} XRBs, and {359} CVs, the optimized classification (i.e., the one using the best set of coefficients as described in Sect. 3.4) 
returns {19813} classifications as AGN, {4702} as stars, {501} as XRBs, and {144} as CVs; {97.7\%} of these classifications agree with the source class ({96.9\%} when computed on a sample with the same class proportions as the priors), although this percentage varies a lot with class. To visualize the classification results in greater detail, we compute the confusion matrix and the metrics introduced before: Table 4 shows that the classification performs best on AGN and stars, reasonable for XRBs and quite poor for CVs, giving lower retrieval fractions and higher false positive rates. The $precision$ rates in Table 4 are computed using Eq. (5).\\


From Table 4, we get the average $f_1$ score of 0.841. This is considerably better than the two naive baselines: a random-guess classifier would have an average $f_1$ score of 0.190 and a classifier that always chooses the largest class (AGN), 0.199.

\begin{table}[]
    \centering
     \caption{Confusion matrix and metrics of the classification applied to the reference sample of 2SXPS.}
    \begin{tabular}{c|cccc|c}
        Truth $\rightarrow$ & AGN & Star & XRB & CV & Total cl.\\\hline
        $\rightarrow$AGN & {19515} & {82} & {25} & {191} & {19813}\\
        $\rightarrow$Star & {44} & {4628} & {3} & {27} & {4702} \\
        $\rightarrow$XRB & {140} & {18} & {326} & {17} & {501}\\
        $\rightarrow$CV & {9} & {9} & {2} & {124} & {144} \\
        \multirow{1}{*}{Total} & {19708} & {4737} & {356} & {359} & \multirow{1}{*}{Average}\\\hline 
        $recall$ (\%) & 99.0 & 97.7 & 91.6 & 34.5 & 80.7\\
        $precision$ (\%) & 97.0 & 98.6 & 90.7 & 85.5 & 92.3\\
        $f_1$ score & .980 & .981 & .911 & .492 & .841 \\
    \end{tabular}
   \tablefoot{ 
    The first four rows show the number counts of objects classified as AGN, stars, XRBs, and CVs, respectively. 
    The last three rows give the $recall$ rate, the true positive rate ($precision$), and the $f_1$ score associated with each class, and those averaged over all classes are given in the last column.}
    \label{tab:4}
\end{table}

Unlike a random forest method (Sect. 5.2), our model gives insight into the motives of each classification. This asset we call ``classification interpretability'' will be used in the following to constrain as much as possible the classification performance and biases.

To begin with, we analyzed the population of XRB false positives, and about 30 of them by manual inspection. These {175} 
sources wrongly classified as XRBs mainly consist of 
AGN (140), 
the rest being CVs (17) and stars (18). We note that {29}\% of them are actually misclassified by a narrow margin (there is less than 20\% difference between the XRB probability and the probability of the correct class). Then after manual inspection, it appears that the reasons for misclassification as XRBs are diverse: {primarily (in 60\% of them), a peculiar spectrum (either too soft or too hard, or a computed luminosity in the XRB range) played an important role. About 30\% presented a rather unexpected variability (typical flux ratio of 10), 
hence mimicking the variability of many XRBs}.
Figure 4 shows illustrations of these reasons for two typical false positives: for the source {2SXPS J095636.3+690027, which is associated with SDSS J09566+6900, a well-known quasar from \cite{Veron2010}, we see that the X-ray luminosities are responsible for the high XRB probability given by the classifier. Because it is located 40 arcsec away from the center of M81, it was wrongly associated with this galaxy and thus its first X-ray luminosity ($1.2\times 10^{38}$ erg~s$^{-1}$) was wrongly computed. The second X-ray luminosity ($1.5\times 10^{32}$ erg~s$^{-1}$) is wrong as well, since it stems from a distance of 4 kpc given by the Gaia counterpart, as put in the catalog of stars of \cite{GaiaDist2021}}. Regarding 2SXPS J125801.1+013431, the probability "track" (left panel) shows that its variability properties are responsible for its classification as an XRB, which is confirmed by its light curve (top right panel): Its flux varied by more than two orders of magnitude over 6 years, which is explained by the tidal disruption of a super-Jupiter \citep{Nikolajuk2013}. Some other XRB false positives consist of unusual objects (like AGN of low luminosity (in particular Seyfert-2) in close galaxies where the stellar bulge dominates optical emission, or absorbed distant AGN). 


\begin{figure}
    \centering
    \includegraphics[width=8cm]{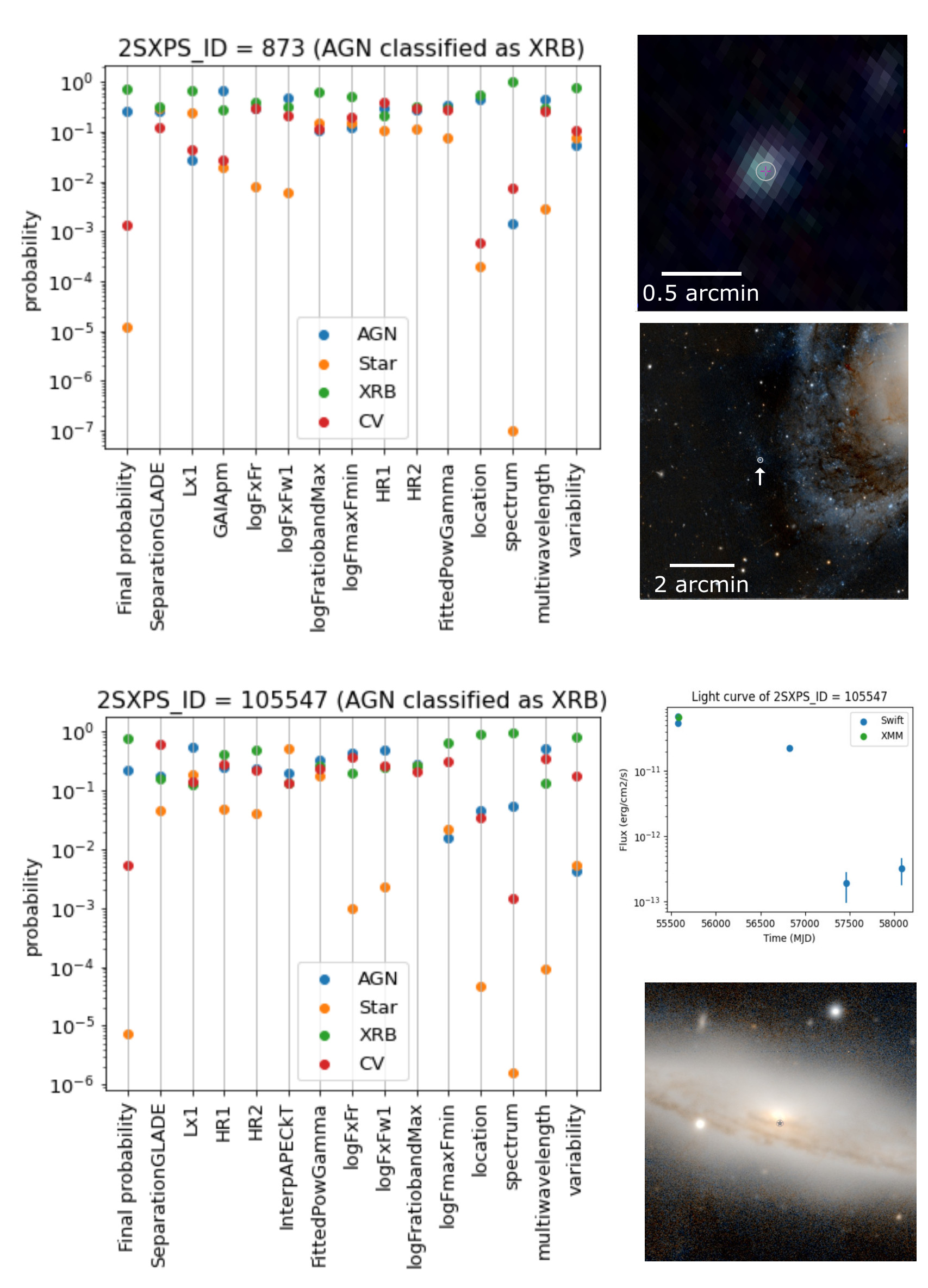}
    \caption{Two examples of XRB false positives. {(Top)} {2SXPS J095636.3+690027, a known AGN located a few arcminutes from M81}. The left panel shows the probability "track", i.e., the likelihoods of each class computed as a function of the properties of the source and then used by the classifier to compute the posterior probabilities of each class (Eqs. (2) and (3)). The top-right panel shows the XMM-EPIC image of the region, the \textit{Swift}-XRT source location, and its error circle (reticle and gray circle). The bottom-right panel shows the Pan-STARRS image of the region and the \textit{Swift} source location. {(Bottom)} 2SXPS J125801.1+013431, the central X-ray source of NGC 4845, known to host an AGN. The left panel shows the probability track, the top-right panel shows the X-ray light curve from the \textit{Swift} and \textit{XMM} detections, and the bottom-right panel shows the Pan-STARRS image of the galaxy and the location of the X-ray source (gray circle).}
    \label{fig:4}
\end{figure}

We also analyzed in each class the sources not retrieved in their class by the classification, and about a hundred of them were manually inspected. For example, {30} XRBs are missed by the classification. They consist of 25 objects classified as AGN, 3 as stars, and 2 as CVs. Many of them (35--40\%) were simply not variable enough during the X-ray detections to be classified as XRBs. Similarly, 35\% have a low classification margin and would be classified as XRBs if the prior value of XRBs (7\%) was somewhat higher. {Another $\sim$25\% actually have dubious identifications, and the catalogs used to form the training sample may not be trusted for these objects (e.g., some X-ray sources seemingly hosted by a galaxy and hence considered as XRBs, but also present in a catalog of quasars).} 
The rest (5-10\%) are generally extragalactic XRBs in a globular cluster for which the multiwavelength counterpart is the whole cluster.


Likewise, {109} stars are wrongly classified as AGN (82), XRBs (18), {or CVs (9)}. They most often consist of particularly variable types of star (eclipsing binaries, young stellar objects, etc.), X-ray sources harder than typical stars or {sources with a very low proper motion (few mas/yr). A few of them ($\sim$10\%) are actually not stars, but HMXBs or distant Seyfert~1 galaxies according to Simbad.}
Regarding the 235 CVs not retrieved by the classification, the misclassification seems often to be due to a lack of follow-up by X-ray observatories, or the fact that we did not characterize the light curves of single observations, so that little to no variability is detected. In particular novae are often classified as AGN (in about {60\%} of cases) because their X-ray hardness and multiwavelength profile are compatible with this type, which is very favored by the priors. Some types of CVs are more easily retrieved, such as polar CVs (AM Her, more than {75\%} retrieval). 
Still, about {85\%} of missed CVs would have been classified as CVs if equal priors had been applied.


\subsection{Results on the test samples}

The same classification model was applied to the 2SXPS test sample, composed of the $\sim$113000 unknown sources following two or more of the quality rules defined in Sect. 2.1.3. Table 6 shows a random subset of these classifications. As a result we obtain {93000} objects classified as AGN (82\%), {16000} as stars (14\%), {4300} as XRBs (4\%), and {321} as CVs (0.3\%), which is globally consistent with the priors, with some discrepancies for the stellar {and CV} types. 
The first one may be explained by an overestimation of the prior proportion of stars, while the low number of objects classified as CVs in the test sample can be explained by the CV prior being low and the reference sample being heterogeneous for this type. But also, since the classification was not optimized on this type, its retrieval is not maximized.


As expected, these proportions change with the number of quality rules followed by the sources: on the (inappropriate) test sample of unknown sources following one rule, 94\% of them are classified as AGN. However, the proportions of each output class among sources following at least two quality rules seem to be rather independent of the number of followed rules.

In order to assess the results of the classifier on the test sample, we did three different studies: a manual inspection of a sample of 120 unknown sources, an analysis of the property distribution of each output class, and a comparison between the \textit{Swift} and \textit{XMM} classifications.

We examined a randomly selected sample of 205 sources present in both \textit{Swift} and \textit{XMM} catalogs. This was done for simplicity, in order to compare both \textit{Swift} and \textit{XMM} classifications to the manual identification; however, this may introduce a bias toward often-pointed fields, in particular nearby galaxies. For each source, the multiwavelength images, X-ray spectrum, X-ray light curves (both within an observation and between observations) and potential Simbad identification were carefully inspected to infer a classification. We identified {134} AGN, {33} stars, {29} XRBs, {1} CV (confirmed by Simbad), and {8} objects of other nature ({2} pulsars, {5} young stellar objects, and {1} supernova remnant, according to Simbad). 
Then we classified them a second time using the classifier trained on the \textit{Swift} reference sample, and a third time using the classifier trained on \textit{XMM}.

Overall, among the {197} objects identified in one of the 4 classes, {86}\% of \textit{Swift} classifications and {84}\% of \textit{XMM} classifications agree with the manual analysis. In greater detail, the \textit{Swift} classifier retrieves {94}\% of AGN, {76}\% of stars, {61}\% of XRBs, and the CV, whereas the \textit{XMM} classifier retrieves {91}\% of AGN, {73}\% of stars, {64}\% of XRB, and also the CV. {The false positive rates of AGN, stars and XRBs, are about 10\%, 5\%, and 30\%. The \textit{Swift} and \textit{XMM} classifiers agree in {87}\% of the cases, and when the result differs it is in most cases due to a low or unreliable X-ray flux in one of the catalogs}. An example of an XRB candidate given by all these classification methods, and unknown in the literature, is shown in Fig. 5. 2SXPS J151604.0+561614 (4XMM J151604.0+561615) has a flux of about $2\times 10^{-14}$ erg~s$^{-1}$~cm$^{-2}$ and a computed probability to be an XRB of about 98\%, because it has no detected counterpart in optical or infrared and a pretty high variability (it varied by a factor of $\sim$6 in 2 weeks). Its mean observed X-ray luminosity (0.3--10 keV) is about $3\times 10^{38}$ erg~s$^{-1}$, which is well in the XRB range.

\begin{figure}
    \centering
    \includegraphics[width=8cm]{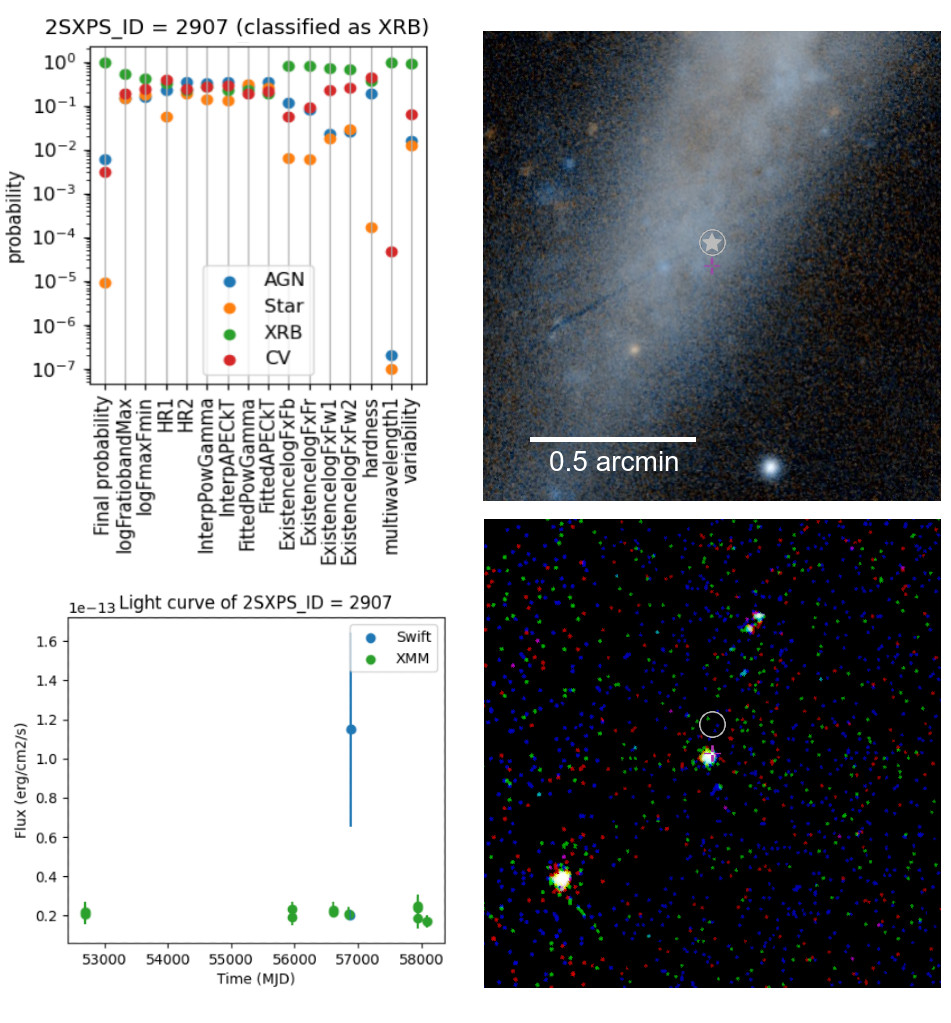}
    \caption{2SXPS J151604.0+561614, an XRB candidate found in NGC 5907 by the classifier. Top left: Probability track showing the likelihood of each class as a function of the source property. Bottom left: \textit{Swift} X-ray light curve (0.3-10~keV) and \textit{XMM} (0.2-12~keV) observations. Bottom right: X-ray image from Chandra; the \textit{Swift} location and position error are pinpointed by the gray circle. Top right: Optical image from Pan-STARRS; the \textit{Swift} location and position error are pinpointed by the circled star.}
    \label{fig:5}
\end{figure}

Another way to verify the self-consistency of the classification model is to statistically compare the populations of each class in the reference sample to their twin as given by the classification. Figure 6 shows such a comparison applied to four source properties. While all these properties show a good overall agreement, especially $HR1$ and $logFratioSnap$, some discrepancies are visible in the two top panels: There is a peak of sources classified as AGN near $b=0^\circ$, and the star and CV distributions of the X-ray to optical flux ratio are shifted to higher values. 
One can interpret these discrepancies as the result of two factors: biases in the reference sample (selection effects), and biases in the classification process. Indeed, the peak of $\log_{10}(F_X/F_r)$ at very low values for reference sample stars is probably a selection effect of nearby stars, which are all well cataloged, while more distant stars at these $\log_{10}(F_X/F_r)$ fall below the sensitivity limit of the X-ray telescope (and thus are not detected). Likewise, AGN behind the galactic plane are under-represented in the literature (e.g., \citealt{Truebenbach2017}), and therefore also in the reference sample: this explains part of the $0^\circ$ peak of sources classified as AGN, being actual AGN. After inspection of Simbad counterparts (about 10\% of this AGN peak), the rest seem to mainly consist of outliers (young stellar objects, stars that are harder than typical stars of the reference sample) and XRBs and CVs with properties in the AGN range and misclassified because of the priors. Last but not least, the shifted $\log_{10}(F_X/F_r)$ peak of CVs is probably the very result of these priors since sources in the bulk of the original peak have much more of a chance of being classified as AGN.

\begin{figure}
    \centering
    \includegraphics[width=8cm]{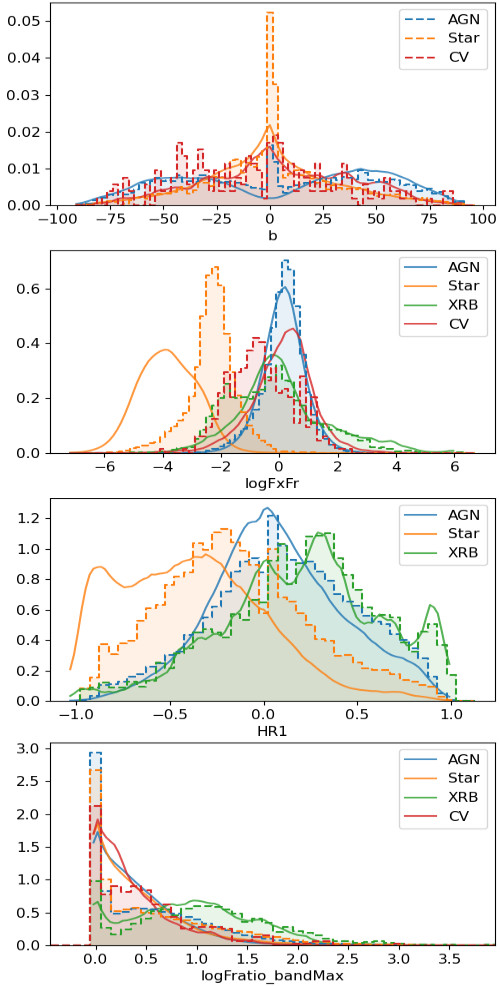}
    \caption{Comparisons between the distributions of different properties in the test sample and their KDE (solid line). From top to bottom: galactic latitude ($b$), X-ray to r-band flux ratio ($\log_{10}(F_X/F_r)$), hardness ratio between soft and medium X-ray bands ($HR1$), and logarithm of the max over mean flux ratio in the most variable XRT band ($logFratioSnap$). The y axis is in arbitrary units. In the first and third panels, only three classes are shown, to ease the graph readability. The $b$ distribution of XRBs resembles a very sharp peak at 0$^\circ$ and some tails that look like the CV distribution. The $HR1$ distribution of CVs shows a good agreement with the KDE, which is shown in Fig. 3.}
    \label{fig:6}
\end{figure}

We also compared the results of \textit{Swift} and \textit{XMM} classifiers to a sample of X-ray sources of good quality and present in both catalogs. This sample consisted of \textit{Swift}-\textit{XMM} matching sources, following at least two quality rules in each catalog, and having reasonable quality flags -- \textit{Swift}: $detflag\le 1, fieldflag\le 1$, \textit{XMM}: $SC\_SUM\_FLAG\le 1$ -- 
({16000} sources at this stage), with the additional constraint of having the same optical and infrared counterparts 
({12200} sources). Depending on the catalog under consideration, the classifier returned $\sim$
{9100} AGN, 
{2100} stars, 
{850} (\textit{XMM}) to {1020} (\textit{Swift}) XRBs, and 
about
{75} CVs. As a result, the two classifiers agree in 
92\% of cases, in particular on AGN classifications 
(96\%), then stars 
(91\%), then XRBs 
(62 to 75\%), and finally CVs 
(40 to 45\%). The causes of differing classifications are diverse, but we note the following trends: They can occur when the candidates have ambiguous classifications ($\sim$ {60}\% of such cases have a final probability of the selected class lower than {65}\%), when mean X-ray fluxes highly differ between \textit{Swift} and \textit{XMM} (factor of $>$ 3 difference in $\sim 33$\% of sources), and when two categories suggest two different classifications. In the last case the difference is thus due to the two versions of weighting coefficients resulting from the optimization, leading the \textit{XMM} classifier to a higher XRB retrieval fraction at the expense of the XRB false positive rate.

\subsection{Classification of 4XMM-DR10 sources}

We wanted to determine how the classification efficiency evolves between two different X-ray surveys. To answer this question, we performed the same parameter optimization for 4XMM-DR10 as for 2SXPS, this resulted in the following output coefficients: {7.5} for location, {3.2} for hardness and luminosity, {2} for multiwavelength ratios, and {6} for variability (Table 3). Eventually, the metrics and confusion matrix of the classification when it was performed on the training sample are given in Table 5. The result is that our model gives consistent results for about {97}\% of sources. However, just as for 2SXPS, classifications as CV must be taken with caution. The average $f_1$ score of this classifier is {very similar to} the one of 2SXPS.

\begin{table}[]
    \centering
    \caption{Same as Table 4, but for the reference sample of 4XMM-DR10.}
    \begin{tabular}{c|cccc|c}
        Truth $\rightarrow$ & AGN & Star & XRB & CV & Total cl.\\\hline
        $\rightarrow$AGN & {18373} & {25} & {46} & {149} & {18593}\\
        $\rightarrow$Star & {15} & {6197} & {10} & {12} & {6234} \\
        $\rightarrow$XRB & {80} & {12} & {479} & {10} & {581}\\
        $\rightarrow$CV & {4} & {0} & {8} & {81} & {93} \\
        \multirow{1}{*}{Total} & {18472} & {6234} & {543} & {252} & \multirow{1}{*}{Average}\\\hline 
        $recall$ (\%) & 99.5 & 99.4 & 88.2 & 32.1 & 79.8\\
        $precision$ (\%) & 97.2 & 98.9 & 93.7 & 84.6 & 93.6\\
        $f_1$ score & .983 & .991 & .909 & .466 & .837\\
    \end{tabular}
    \label{tab:5}
\end{table}

{A manual analysis of 30 XRB false positives showed that the majority of them (83\%) overlap the extent of a galaxy, and hence have wrong values of X-ray luminosity and a misleading location, and 40\% showed an unexpected multi-instrument variability -- either physical or due to a difference in flux computation method (spectral fitting for bright Swift sources, fixed spectrum for XMM sources). Therefore, sources wrongly classified as XRBs are not misclassified for the same reasons as in 2SXPS: Fewer are primarily misclassified due to an unexpected variability, whereas more misclassifications coincide with the proximity of a galaxy. This may be due to the fact that galaxies are more often targeted by \textit{XMM} pointings than by \textit{Swift} pointings.}

A sample of so-called XRBs un-retrieved by the classification was also analyzed, showing results comparable with the study on \textit{Swift}: un-retrieved XRBs are primarily missed when they vary little between observations or when their multiwavelength ratios are consistent with an AGN. Still, {about 65\% of the sample would have been classified as XRBs if the XRB prior was somewhat higher}. A substantial fraction (about 15\%) of the sample was found to be dubious XRBs (i.e., background AGN or foreground stars contaminating the XRB reference sample).

Once applied to the 4XMM test sample of {345,000} sources, the classifier gives $\sim${270000} AGN (78\%), $\sim$66000 stars (19\%), $\sim$8200 XRBs (2.4\%), and $\sim$1700 CVs (0.5\%). Table 7 shows a random subset of these classifications. These proportions are in good agreement with those obtained on the \textit{Swift} test sample.

A closer look at sources classified as XRBs in the test sample confirmed {45}\% of them as being reliable and {30\% as false positives with misleading data (e.g., variable AGN or AGN in the background of a galaxy)}; another {10}\%  have data quality issues (e.g., {a wrong optical counterpart, a spurious variability in the multi-instrument light curve or} an X-ray source being actually the confusion of two distinct sources), and $\sim$15\% are objects of another type (not AGN, stars, XRBs, or CVs, {in particular young stellar objects}). As expected, sources classified as AGN and stars in the test sample are more reliable: about 95\% {and 99\%} of the ones analyzed manually (about 200) prove to be reliable candidates.



 \subsection{Application: Decontaminating HLX candidates}

An interesting application of classifying X-ray sources is to find candidates of rare types, and to find them with low false positive rates. The search for HLX candidates is a typical example of such studies. HLXs are extragalactic X-ray sources with an X-ray luminosity greater than $10^{41}$ erg~s$^{-1}$ and located outside the nucleus of their host galaxy. At such luminosities, the most conservative Eddington limit should imply a black hole mass greater than $10^2$ M$_\odot$, unlike ULXs (off-center X-ray sources with X-ray luminosities $10^{39}$ < $L_X$ < $10^{41}$ erg~s$^{-1}$) which have been shown to be stellar-mass objects for the majority, accreting above the Eddington limit (e.g., \citealt{Bachetti2014}). Hyperluminous X-ray sources are thus of special interest in the search and the study of IMBHs (\citealt{Greene2020}, and see also one of the best IMBH candidate known to date, HLX-1, \citealt{Farrell2009}). They may resemble a scaled-up version of black hole XRBs, thus probably favoring a classification as XRBs. However, samples of ULX and HLX candidates are highly contaminated by foreground stars and background AGN, reaching 70\% of contaminants \citep{Zolotukhin2016}. The decontamination is then done manually. In the following we apply our automatic classifier to a sample of HLX candidates. We thus expect to classify background contaminants as AGN and foreground contaminants as stars.

We built our sample from a positional crossmatch between the \textit{Swift} test sample and the GLADE \citep{Dalya2016} catalog. This sample of X-ray sources matching GLADE galaxies contains $\sim$8000 objects, including $\sim$3000 sources matching the galaxy center. We computed their mean and peak unabsorbed X-ray luminosity in the total band (0.3-10 keV), using their unabsorbed mean (peak) XRT flux and the distance given by GLADE. Figure 7 shows the distribution of the peak unabsorbed X-ray luminosity in our sample. The peak around $10^{33}$ erg~s$^{-1}$ consists of sources in globular clusters of the Milky Way. At higher luminosities the bimodal distribution seems to be the sum of the population of off-center sources (peak around $10^{39}$ erg~s$^{-1}$) and the one of central sources (peak around $10^{43}$ erg~s$^{-1}$, which is compatible with typical AGN luminosities). Off-center XRB candidates seem to fill a quite narrow range of luminosities, between $10^{37}$ and $10^{42}$ erg~s$^{-1}$.

\begin{figure}
    \centering
    \includegraphics[width=8cm]{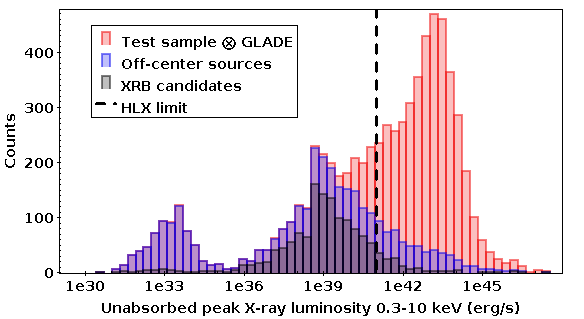}
    \caption{{Distribution of the X-ray peak luminosity} of \textit{Swift} sources matching a galaxy from GLADE. Only sources with at least two quality rules and reasonable quality flags are selected. The blue bars show the subset of off-center sources, i.e., sources whose error circle does not match the galaxy center. The dark gray bars show a subset of these sources, made of good XRB candidates (candidates with $\mathbb{P}(XRB)>2\mathbb{P}(c)$ for every other class $c$).}
    \label{fig:7}
\end{figure}

The classifier returns different results depending on the center or off-center location of the source: sources matching the center are most often classified as AGN {(88\%), then as XRBs (11\%, largely because of a very bright bulge of stars dominating the optical emission and a high variability)}. On the other hand, off-center sources are less often classified as AGN ({42}\%), more often as XRBs ({52}\%), and then as stars ({5}\%): At first sight, this sample seems to be contaminated by $\sim2/3$ of sources not belonging to their matching "host". When selecting only off-center sources in the HLX luminosity range, this assessment is even clearer: these HLX candidates consist of {71}\% of sources classified as AGN, {27}\% (163) as XRBs, and {2}\% as stars. We are thus left with {163} "decontaminated" HLX candidates, and {129} of them seem to be "strong" candidates (having $\mathbb{P}(XRB)>2\mathbb{P}(\mathrm{c})$ for every other class c). These ones notably include {six} known ULXs and HLX-1, but also 80 seemingly unknown sources (no entry in Simbad). They will be studied in greater detail in a future work.

 \subsection{Outliers}
 
By construction, the classifier we built has to choose a class for every object from a small selection (here AGN, stars, XRBs, and CVs). Some X-ray sources do not belong to any of these classes (e.g., spurious sources, X-ray transients, galaxy clusters, supernova remnants) and their classification will thus be misleading. However, some of these outliers may be retrieved by analyzing 
their outlier measure (as defined in Sect. 3.1).

This approach has already been carried out in the work of \cite{Lo2014}: they considered a source as being anomalous if its classification margin (i.e., $2\mathbb{P}(\mathrm{c}|data)-1$ where c is the preferred class and $\mathbb{P}$ is the output probability) was less than $-0.3$, or if its outlier measure - not the same as the one we used -  was greater than 10. As a result they obtain mainly sources with data quality issues, but also an interesting HLX candidate.

Figure 8 shows the distribution of the outlier measure for different subsets of 4XMM-DR10 sources. AGN false positives (in red in the first panel) tend to have a higher outlier measure than actual AGN, due to some of their properties differing significantly from the bulk of AGN in the reference sample. Such objects are generally classified as AGN by default, because of the higher prior of this class ({91}\% of them follow $\mathbb{P}(AGN)<0.9$, against 5\% of actual AGN). Some AGN from the reference sample also have high outlier measures, because their properties also deviate from the bulk of reference AGN: Indeed, as obtained after a cross-correlation with Simbad, more than 70\% of them are actually quasi-stellar objects, and less than 13\% and 1\% are respectively Seyfert 2s and LINERs, which are types of AGN whose nuclear emission is absorbed by the dusty torus, while the distribution of $\log_{10}(F_X/F_r)$ is centered on 0 for quasars and -2 for LINERs (Seyfert 2 also shine brighter in optical and are much harder in X-ray due to absorption). The same conclusions can be drawn for the stellar type, with variable stars that have the highest outlier measures. Non-stars classified as star generally have some properties deviating from typical stars (either the hardness or the variability). The outlier measure can then be used to retrieve a substantial fraction of objects not matching one of the four classes we used.

To better characterize the outliers, we studied a sample of 85 objects that have an outlier measure higher than 10 and that match a Simbad source. Their analysis can be divided by their Simbad type, between AGN, stars, XRBs, and other types. {About 55\%} of AGN outliers are just LINERs, Seyfert 2, and blazars, in which the outlier has a physical meaning. Another {10\%} of them are outliers because of their variability: these AGN vary in X-ray by a factor of 5 to 100 between observations, sometimes due to a faint flux (the variation is contained in the error bars) but most often well-founded when inspecting the light curve); {the rest of outliers is due to the location of the source, for example when it is an AGN in the background of a galaxy, or a peculiar hardness in X-rays}. Similarly, {two thirds} of star outliers are just young stellar objects and variable stars of Orion-type, and the rest of outliers are due to a large X-ray-to-optical flux ratio and/or a significant X-ray variability. X-ray binaries are numerous to have high outlier measures, due to the heterogeneity of this class, but those with a measure higher than 10 are generally {those with no infrared counterpart or no counterpart at all, or with a particularly hard spectrum, or located in the Magellanic clouds.}

\begin{sidewaystable}[]
  \centering\tiny
  \caption{Classification of some 2SXPS sources from the test sample.}
  \begin{tabular}{c|ccccccccccccccccccccccc}
IAUNAME &$Err_{1\sigma}$        &Flux   &det    &Q      &b      &pm     &$L_X$  &HR1    &HR2    &FxFw1  &dIR    &Fratio &$P_{AGN}$      &$P_{Star}$     &$P_{XRB}$      &$P_{CV}$       &cl     &alt    &O.M.\\
$^{(1)}$        &$^{(2)}$       &$^{(3)}$       &$^{(4)}$       &$^{(5)}$       &$^{(6)}$       &$^{(7)}$       &$^{(8)}$       &$^{(9)}$       &$^{(10)}$      &$^{(11)}$      &$^{(12)}$      &$^{(13)}$      &$^{(14)}$      &$^{(15)}$      &$^{(16)}$      &$^{(17)}$      &$^{(18)}$      &$^{(19)}$      &$^{(20)}$\\
J002927.8+091434        &2.19   &3.7e-12        &14     &3      &-53.2  &         &       &0.04   &-0.22  &0.78   &1.39   &2.48   &0.17   &0.0    &0.78   &0.04   &XRB         &AGN            &8.94\\
J020142.4-290954        &3.12   &2.6e-14        &1      &2      &-74.4  &         &       &0.62   &-0.31  &-1.27  &3.49   &0.0    &0.99   &0.0    &0.01   &0.01   &AGN         &               &6.52\\
J065241.1+400658        &2.69   &2.4e-14        &2      &3      &17.3   &0.54   &         &-0.34  &0.83   &-1.64  &6.24   &0.17   &1.0    &0.0    &0.0    &0.0    &AGN         &               &6.49\\
J080044.1-563851        &2.5    &1.5e-13        &1      &2      &-13.6  &2.48   &         &-0.01  &0.5    &-2.02  &4.91   &0.0    &0.95   &0.01   &0.01   &0.03   &AGN         &               &8.13\\
J080156.5+770656        &3.0    &1.6e-13        &1      &2      &30.3   &         &       &0.79   &-0.01  &-1.78  &9.56   &0.0    &0.98   &0.0    &0.01   &0.01   &AGN         &               &6.84\\
J085002.5+521521        &3.06   &7.7e-15        &1      &2      &39.1   &0.44   &         &-0.54  &0.83   &-2.64  &4.86   &0.0    &0.98   &0.02   &0.0    &0.0    &AGN         &               &7.48\\
J091450.7+440035        &2.62   &5.1e-14        &1      &2      &43.8   &37.37  &         &0.3    &-0.82  &-2.17  &5.36   &0.0    &0.03   &0.93   &0.01   &0.03   &Star   &                 &8.17\\
J095851.9-284208        &3.44   &3.8e-14        &1      &2      &20.5   &1.07   &         &0.21   &-0.08  &-1.63  &2.89   &0.0    &0.93   &0.0    &0.02   &0.04   &AGN         &               &7.61\\
J101543.2-334048        &3.38   &3.9e-14        &2      &3      &18.9   &         &       &0.05   &-0.26  &-1.14  &2.5    &0.25   &0.98   &0.0    &0.01   &0.01   &AGN         &               &5.35\\
J123149.2+110654        &2.87   &5.4e-14        &2      &3      &73.3   &         &       &0.07   &-0.1   &-1.57  &4.37   &0.18   &0.98   &0.0    &0.01   &0.01   &AGN         &               &4.7\\
J123610.0-643500        &1.69   &4.1e-14        &1      &2      &-1.8   &2.5    &         &0.8    &0.44   &-1.88  &2.95   &0.0    &0.92   &0.0    &0.03   &0.04   &AGN         &               &8.28\\
J150957.3-193041        &2.19   &1.5e-13        &1      &3      &32.5   &         &43.48  &0.2    &-0.29  &-0.97  &4.67   &0.0    &0.99   &0.0    &0.01   &0.0    &AGN         &               &9.87\\
J160205.9+663619        &2.87   &6.7e-14        &1      &2      &41.0   &         &       &-0.08  &0.17   &-1.56  &5.01   &0.0    &0.99   &0.0    &0.0    &0.01   &AGN         &               &4.97\\
J173317.5+491910        &2.69   &1.8e-14        &2      &3      &32.7   &         &       &-0.34  &0.34   &-1.22  &6.77   &0.15   &0.98   &0.0    &0.01   &0.0    &AGN         &               &5.42\\
J205048.8-254443        &4.19   &6.5e-14        &1      &2      &-36.7  &         &       &-0.42  &0.56   &-1.63  &5.66   &0.0    &0.97   &0.01   &0.02   &0.01   &AGN         &               &6.02\\
  \end{tabular}
 \tablefoot{ The full table is available at \url{https://github.com/htranin/classificationXray/raw/main/data/2SXPSout.fits} and on VizieR. Description: $^{(1)}$ IAUNAME of the source (without prefix ``2SXPS'') $^{(2)}$ $1\sigma$ \textit{Swift} position error (arcsec) $^{(3)}$ Mean total band flux (0.3--10keV) of all \textit{Swift} detections of this source $^{(4)}$ Number of X-ray detections, including \textit{XMM} and \textit{Chandra} $^{(5)}$ Number of quality rules followed $^{(6)}$ Galactic latitude, in degrees $^{(7)}$ Gaia proper motion of the optical counterpart $^{(8)}$ Logarithm of the mean X-ray luminosity, computed using the distance of associated GLADE galaxy $^{(9, 10)}$ Hardness ratios between the bands 0.3--1 and 1--2 keV (resp. 1--2 and 2--10 keV) of \textit{Swift} $^{(11)}$ Logarithm of the X-ray to W1-band flux ratio  $^{(12)}$ Angular distance (arcsec) to the infrared counterpart  $^{(13)}$ Logarithm of the maximum to minimum X-ray flux ratio  $^{(14-17)}$ Probabilities of each class  $^{(18)}$ Class giving the highest probability  $^{(19)}$ Alternative classification, if any, when a property category is ignored  $^{(20)}$ Outlier measure (Sect. 3.2).}
 \label{tab:6}
\end{sidewaystable}

\begin{sidewaystable}[]
  \centering\tiny
  \caption{Classification of some 4XMM-DR10 sources from the test sample.}
  \begin{tabular}{c|ccccccccccccccccccccccc}
IAUNAME &$Err_{1\sigma}$        &Flux   &det    &Q      &b      &pm     &$L_X$  &FxFw1  &dIR    &HR1    &HR2    &Fratio &$P_{AGN}$      &$P_{Star}$     &$P_{XRB}$      &$P_{CV}$       &cl     &alt    &O.M.\\
$^{(1)}$        &$^{(2)}$       &$^{(3)}$       &$^{(4)}$       &$^{(5)}$       &$^{(6)}$       &$^{(7)}$       &$^{(8)}$       &$^{(9)}$       &$^{(10)}$      &$^{(11)}$      &$^{(12)}$      &$^{(13)}$      &$^{(14)}$      &$^{(15)}$      &$^{(16)}$      &$^{(17)}$      &$^{(18)}$      &$^{(19)}$      &$^{(20)}$\\
J004238.9+401630        &1.72   &1.1e-14        &1      &2      &-22.6  &6.99   &35.9   &-0.46  &3.58   &-0.04  &-0.09  &0.0    &0.56   &0.09   &0.24   &0.12   &AGN         &Star XRB       &10.86\\
J053031.7-265616        &0.78   &1.6e-13        &1      &3      &-28.7  &0.13   &         &0.19   &0.68   &0.11   &-0.04  &0.0    &1.0    &0.0    &0.0    &0.0    &AGN         &               &6.48\\
J074147.6+702203        &2.56   &9.5e-14        &1      &2      &29.6   &0.44   &         &0.82   &1.49   &-0.73  &0.86   &0.0    &1.0    &0.0    &0.0    &0.0    &AGN         &               &7.29\\
J080142.9+004342        &1.71   &1.9e-14        &1      &2      &15.9   &3.0    &         &-0.14  &2.24   &0.36   &0.47   &0.0    &0.93   &0.0    &0.01   &0.05   &AGN         &               &7.58\\
J090143.2+355133        &0.5    &8.5e-14        &1      &3      &41.0   &107.89 &         &-3.46  &1.25   &0.52   &-0.57  &0.0    &0.0    &1.0    &0.0    &0.0    &Star   &                 &7.46\\
J104121.8+395616        &0.94   &1.1e-14        &2      &2      &60.2   &         &       &-0.17  &0.97   &0.91   &0.54   &0.07   &0.98   &0.0    &0.02   &0.01   &AGN         &               &6.34\\
J123536.6+123059        &1.18   &1.3e-14        &8      &4      &74.9   &         &38.59  &-0.17  &2.89   &-0.07  &0.51   &0.57   &0.05   &0.0    &0.95   &0.0    &XRB         &AGN            &8.45\\
J124649.0-593539        &0.81   &5.5e-15        &1      &3      &3.3    &5.5    &         &-3.56  &1.17   &0.79   &-0.15  &0.0    &0.0    &1.0    &0.0    &0.0    &Star   &                 &6.93\\
J165142.5-385515        &2.23   &1.4e-14        &1      &2      &3.4    &11.76  &         &-2.61  &0.74   &0.91   &-0.12  &0.0    &0.0    &1.0    &0.0    &0.0    &Star   &                 &7.18\\
J174739.0-282139        &1.28   &1e-14  &1      &2      &-0.1   &2.22   &         &-1.35  &3.66   &-0.01  &-0.41  &0.0    &0.58   &0.32   &0.04   &0.06   &AGN         &Star           &7.75\\
J182009.4-155744        &2.15   &2.2e-13        &1      &2      &-0.5   &         &       &-1.99  &3.24   &1.0    &0.26   &0.0    &0.39   &0.5    &0.04   &0.07   &Star   &AGN                 &7.67\\
J191357.0-211606        &0.65   &1.8e-13        &1      &3      &-14.3  &0.17   &         &0.13   &0.95   &0.29   &0.1    &0.0    &1.0    &0.0    &0.0    &0.0    &AGN         &               &6.66\\
J204846.5-095358        &1.16   &5.1e-15        &1      &2      &-30.5  &         &       &-0.58  &2.49   &0.25   &-0.08  &0.0    &0.97   &0.01   &0.01   &0.01   &AGN         &               &6.27\\
J213521.0+312702        &3.09   &1.5e-13        &1      &3      &-15.1  &8.43   &         &1.32   &8.39   &-0.18  &0.59   &0.0    &0.84   &0.01   &0.01   &0.14   &AGN         &               &8.33\\
J221635.7+000902        &1.67   &3.5e-14        &3      &3      &-43.9  &         &       &0.73   &4.24   &0.59   &0.39   &0.54   &0.93   &0.0    &0.05   &0.02   &AGN         &               &6.53\\
  \end{tabular}
  \tablefoot{
   The full table is available at \url{https://github.com/htranin/classificationXray/raw/main/data/XMM10out.fits} and on VizieR. Description: $^{(1)}$ IAUNAME of the source (without prefix ``4XMM'') $^{(2)}$ $1\sigma$ \textit{XMM} position error (arcsec) $^{(3)}$ Mean total band flux (0.2--12keV) of all \textit{XMM} detections of this source $^{(4)}$ Number of X-ray detections, including \textit{Swift} and \textit{Chandra} $^{(5)}$ Number of quality rules followed $^{(6)}$ Galactic latitude, in degrees $^{(7)}$ Gaia proper motion of the optical counterpart $^{(8)}$ Logarithm of the mean X-ray luminosity, computed using the distance of associated GLADE galaxy $^{(9, 10)}$ Hardness ratios between the bands 0.2--0.5 and 0.5--1 keV (resp. 0.5--1 and 1--2 keV) of \textit{XMM} $^{(11)}$ Logarithm of the X-ray to W1-band flux ratio  $^{(12)}$ Angular distance (arcsec) to the infrared counterpart  $^{(13)}$ Logarithm of the maximum to minimum X-ray flux ratio  $^{(14-17)}$ Probabilities of each class  $^{(18)}$ Class giving the highest probability  $^{(19)}$ Alternative classification, if any, when a property category is ignored  $^{(20)}$ Outlier measure (Sect. 3.2).
   \label{tab:7}
}
\end{sidewaystable}

Sources of other types considered as outliers include many transients ({12} gamma-ray bursts, {17} supernovae, and {5} TDEs, 62 supernova remnants, 17 pulsars, and 11 planetary nebulae. Only 40\% of sources of these types have an outlier measure higher than ten, either because the others lack critical data (25\% of transients have only one X-ray detection) or because the properties we use do not enable them to be distinguished from the four types: AGN, stars, XRBs, and CVs.

Last but not least, we analyzed a sample of sources that have an outlier measure higher than ten and no counterpart in Simbad. After a careful inspection of the \textit{XMM} fields (and Chandra, when available), among the {40} sources analyzed, {36\% have a low detection likelihood and may be spurious. The other 64\% are actual sources not present in Simbad, in particular XRB candidates and AGN in the background but also some galaxy clusters}. 

As a conclusion, our outlier measure is useful to nuance the interpretation of a particular classification, and retrieve some misclassified sources; a non-negligible part of sources deviating from the bulk of classes used as reference; and sources that have data quality issues. However, given the generality of the classes we used, we estimate that only $\sim$1\% of the {14000} 4XMM sources considered as outliers are actually sources of another type. A detailed study of outliers of special interest will be addressed in a future work.

\begin{figure}
    \centering
    \includegraphics[width=8cm]{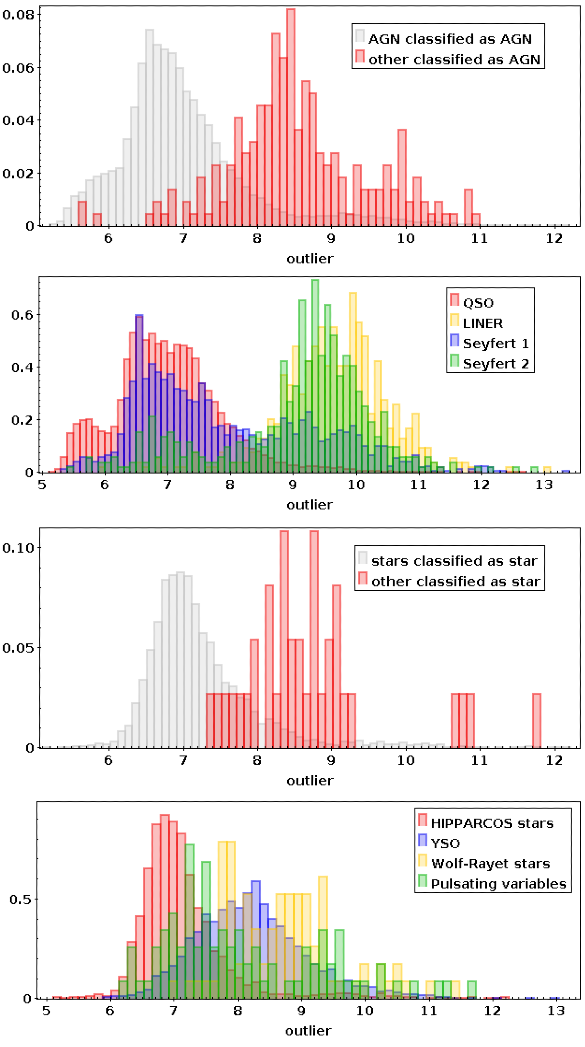}
    \caption{{Distribution of the outlier measure}, obtained from objects in the 4XMM reference sample classified as AGN (first panel), from different types of AGN (second panel), from objects in the reference sample classified as stars (third panel), and from different types of stars (fourth panel).}
    \label{fig:8}
\end{figure}

\section{Discussion}

\subsection{General classification performance}


From the reference sample, we can expect that {at least} 95\% of all objects are classified correctly, in both \textit{Swift} and \textit{XMM} catalogs. However, our results on the test sample show a slightly lower performance, about {85--90}\%. This is expected due to physical differences between the reference and the test samples, biases in the classification process, and computation errors of some properties in the catalog, as will be detailed in the rest of Sect. 5. 

Manual inspection of the test samples still showed an overall good performance of our classification, with about {95}\% of sources classified as AGN and stars actually being reliable candidates. Despite the optimization, XRBs remain less well classified than AGN and stars: this is partly due to their lower prior as well as the wide variety of properties they can have. Sources classified as XRBs are rarer and their reliability is lower, because they include a substantial fraction of outliers that have a particular physical nature and another substantial fraction of sources affected by data issues (spurious variability or wrongly-missing counterparts). Since the AGN and stars in the reference sample were found by crossmatching with catalogs built in the optical or infrared, we also expect more frequent matching issues in the test sample than in the reference sample. Overall, we infer an XRB true positive rate of 40--50\% and 30--40\% in the \textit{XMM} and \textit{Swift} test samples, respectively. The large difference between these rates and the ones computed on the reference samples is {notably due to the presence of sources of other nature and faint sources with data quality issues -- more frequent than in the reference sample}. 


The CV class gives a lower performance, either because a classification as CVs is difficult to infer with the properties we use (one needs either the spectrum or the light curve), or because the classification is not optimized for this class. Overall, mainly because of their low prior, sources classified as CVs have lower classification margins: about 95\% of them have a probability $\mathbb{P}(\mathrm{CV})<60$\%. Further work is needed to better perform at the classification of this rare type.

Differences exist between \textit{Swift} and \textit{XMM} performances, due to a better constrained variability (\textit{Swift}) or better constrained hardness and location (\textit{XMM}). {The X-ray luminosity and variability and the proximity to a galaxy} prove to be the most important features from the optimization, regardless of the catalog. However, this may change if we optimized it for, for example, AGN or CVs.






\subsection{Comparison to machine learning results}

For comparison, we performed a classification of 2SXPS sources using a random forest algorithm with the following parameters: four classes (AGN, star, XRB, CV), the same source properties as presented in Table 1, a maximum depth of the decision trees equal to 12, and a weighted loss function to optimize the classification for XRBs (the weight of this class is set to 1 while the weight of other classes is set to 0.01). The algorithm was trained on a sample containing 70\% of the reference sample and validated on the other 30\%. The confusion matrix obtained on the validation sample is shown in Table 8. Compared to Table 4, we see comparable results. The $precision$ of XRBs and their $recall$ seem to be slightly higher and lower, respectively, but this depends on the weight chosen for the XRB class. The $f_1$ score of XRBs is also {higher than} a recent random forest classifier of X-ray sources in the literature (0.769 for \citealt{Arnason2020}, however not based on the same source properties).

\begin{table}[h]
    \centering
    \caption{Same as Table 4, but for the random forest classification applied to the reference sample of 2SXPS.}
    \begin{tabular}{c|cccc|c}
        Truth $\rightarrow$ & AGN & Star & XRB & CV & Total cl.\\\hline
        $\rightarrow$AGN & {5889} & {7} & {20} & {39} & {5955}\\
        $\rightarrow$Star & {6} & {1404} & {1} & {3} & {1414} \\
        $\rightarrow$XRB & {9} & {5} & {83} & {5} & {102}\\
        $\rightarrow$CV & {7} & {1} & {1} & {68} & {77} \\
        \multirow{1}{*}{Total} & {5911} & {1417} & {105} & {115} & \multirow{1}{*}{Average}\\\hline 
        $recall$ (\%) & 99.6 & 99.1 & 79.0 & 59.1 & 84.2\\
        $precision$ (\%) & 96.8 & 99.2 & 95.2 & 87.9 & 95.2\\
        $f_1$ score & .982 & .991 & .863 & .707 & .886\\
    \end{tabular}
    \label{tab:8}
\end{table}

We compared random forest performance to our classification on the test sample, both statistically and with manual inspection of a small subsample.

Once trained on the whole reference sample and applied to the 2SXPS test sample defined in Sect. 4.2, we obtain about 97000 sources classified as AGN, 13000 as stars, 2200 as XRBs, and 330 as CVs. Compared with the naive Bayes classifier, this classification thus gives similar class proportions, with a little preference for the AGN {and star} classes at the expense of the XRB one. However, the results obtained with the naive Bayes classifier are more interpretable because the motives of each classification are explicitly related to the values of the source properties.

\subsection{Classification biases}

Biases can occur both before and during a classification process, and these two types of bias are sometimes hard to disentangle. For instance, X-ray surveys naturally contain few AGN at low galactic latitudes, so unknown sources at low latitudes are less likely to be classified as AGN than sources at higher latitudes. However, this difference in AGN frequency may not be the same in the training sample and in the whole sample, since the training sample is partly built up using surveys searching explicitly for extragalactic sources, which thus exclude lower latitudes by construction. The proportion of each class in the training sample is thus biased in an unquantifiable way. This is partly corrected by manual choice of the priors, but this still affects the probability densities of each class.

A second type of bias refers to missing values, in particular the multiwavelength flux ratios. The non-detection of multiwavelength counterparts is considered as a feature as well as other properties, so it is important to not miss them because of statistical effects. If such a value is missing for a particular object, this means that no counterpart were found {by NWAY}, and manual inspection revealed that possibly missing counterparts are {rare}. 
However, when a source is in a stellar association, a spurious counterpart is identified sometimes, in particular {the optical counterpart may be different than the infrared one}. More importantly, while missing counterparts for an X-ray bright source give a significant piece of information, a faint source that has no counterparts is not surprising at all, whereas our model does not take the X-ray flux into account. Similarly, a source that has just one detection has the same variability parameters as a source of constant luminosity and is considered as such by the algorithm. 

We expect also important biases for this study when applying a model optimized for the training sample to the test sample, which has not exactly the same properties. In particular, since most X-ray emitting stars have to be close to be detected, many of them are already in the reference sample, which thus has a higher proportion of stars than the test sample. As a consequence, the prior of the stellar class should not be the same between the two samples. Likewise, since the reference sample contains a greater proportion of bright sources, the absence of multiwavelength counterpart may be over-favoring the XRB class, when going from the training to the test sample. Moreover, because nearby, large galaxies are more studied, they are a preferred location for XRBs in the reference sample, which induce a bias in their galactic latitude distribution. If location were considered as important by the classifier, this would have biased the classification as XRBs in the test sample.

\subsubsection{Bias due to the X-ray flux}

When applying the classification to faint sources, the absence of counterpart is wrongly taken as an important piece of information. Indeed, although most X-ray bright AGN have optical and infrared counterparts, this does not hold anymore at faint fluxes: an AGN following $F_X/F_r=5$ (resp. $F_X/F_{W1}=4$) reaches the Pan-STARRS (resp. UnWISE) limiting magnitude when its X-ray flux is about $3-4\times 10^{-14}$ erg~s$^{-1}$~cm$^{-2}$. The fraction of 4XMM sources matching a quasar from SDSS DR14 \citep{Paris2018} (resp. matching an AGN in allWISEagn), and without any infrared (resp. optical) counterpart, is about 10\% in this flux bin, and becomes much higher in fainter bins. This suggests that the classification of sources below this threshold and missing a counterpart may not be completely reliable; these sources represent about {12}\% of the 4XMM test sample ({4}\% for 2SXPS), and about {36}\% (resp. {11}\%) of the sources classified as XRBs.

\subsubsection{Spurious variability}

As explained in Sect. 2.1.1, X-ray detections from different X-ray observatories were combined to augment the catalog with extra variability data. However, this process has several caveats detailed in the following. First, we find some systematics in the comparison of \textit{Swift} fluxes to other observatories: when the source is faint (typically fainter than $5\times 10^{-14}$ erg~s$^{-1}$~cm$^{-2}$), \textit{Swift} fluxes are very often fainter by a factor of two to ten. This may be due to a caveat in the conversion from the count rate to the X-ray flux, since the conversion factor is not computed using the same hypotheses for 2SXPS, 4XMM, and CSC2.

Second, the X-ray telescopes on board \textit{Swift}, \textit{XMM-Newton,} and \textit{Chandra} do not cover the same energy bands: 0.3--10~keV, 0.2--12~keV, and 0.5--7 keV, respectively. The conversion from one band to another require the assumption of a spectral model, which is why we simply kept fluxes as computed in each X-ray catalog. This can lead to a spurious variability in the multi-instrument light curve, especially in very soft sources: for instance among the few hundred stars present in both 2SXPS and 4XMM-DR10, about 70\% have a \textit{Swift} flux higher than the \textit{XMM} flux, and the median ratio between these fluxes is about 1.5.
These numbers become 54\% and 1.05 in the case of AGN.

\subsubsection{Reference sample reliability}

Despite our effort to maximize the purity of the reference sample, it stills contains a small fraction of contaminants. These are for example the presumed mid-infrared-selected AGN that have actually another type (\cite{Secrest2015} state a purity of their sample higher than 95\%), or some presumed XRBs that are actually flaring stars or background AGN \citep[e.g.,][]{Guo2016,Sazonov2017}. Indeed, some of the catalogs we used to identify known XRB estimate a low but non-negligible number of contaminants, especially background AGN (e.g., \citealt{Liu2007}, \citealt{Kundu2007}, \citealt{Mineo2012} from Table 2).

Another issue when building such a training sample is its completeness. Our reference sample is obviously incomplete at faint fluxes, for which few X-ray sources have a known nature. For instance, only 15\% of AGN in the reference sample have an X-ray flux lower than 2 $\times$ 10$^{-14}$ erg~s$^{-1}$~cm$^{-2}$. This selection effect does not apply on stars (55\% of them have a flux lower than 2 $\times$ 10$^{-14}$ erg~s$^{-1}$~cm$^{-2}$), since X-ray visible stars are bright enough in optical or infrared to be identified as such.
Moreover, other selection effects hinder the sample completeness:  too few AGN are identified at low galactic latitudes \citep{Truebenbach2017, Secrest2020}, too few XRBs in further, less-studied galaxies, etc.

\subsubsection{Choice and size of the classes}

The choice of the classes may be crucial to the performance of the classifier: since it is based on the correlation between the source properties and its class, the sharper the features of each property distribution for each class, the more efficient the classifier can be. Any heterogeneity in the composition of a class, like the presence of a small number of objects with very different properties (e.g., LINER-type AGN among the AGN sample, or highly variable stars among all stars), results in a large outlier measure for these objects. On the other hand, a sufficient number of objects in each class must be identified, in order to obtain reliable KDEs. From a visual inspection of the KDE obtained from different 4XMM-DR10 training samples, we estimate that a number of at least 75--100 objects is advisable to infer meaningful KDE.

In order to investigate further how the classifier is sensitive to the choice of classes, we performed the classification on two other reference samples in 4XMM-DR10: one with only two classes (binary classifier), "XRBs" and "non-XRBs", and the other with six classes: AGN, non-variable stars, variable stars and young stellar objects, XRBs in the Milky Way, extragalactic XRBs, and CVs. The binary classifier returned somewhat {less performing} results than the standard classifier for the XRB class: a retrieval fraction of 82.4\% and a $precision$ of 95\%. The resulting $f_1$ score averaged on the two classes is 0.883.

The six-class classifier returned a lower retrieval fraction ({73.3}\%) but higher $precision$ ({95.6}\%) for galactic XRBs, because of the lower prior for this class (only one-seventh of our XRB sample corresponds to galactic XRBs). Gathering back galactic and extragalactic XRBs after the classification has been performed, we retrieve {92.4}\% of them and the $precision$ of XRB classification is {95.3}\%.  Interestingly, other classes slightly benefit from the split in six classes, for example the stars (gathering back young, variable and other stars in the general class of ``star'' after the classification) for which the retrieval fraction becomes {99.7}\% and the $precision$ becomes {99.2}\%. The $f_1$ score averaged over all four general classes becomes {0.852} with this classifier. The detail of the number counts is shown in Table 9.

The optimization of this classifier (favoring the performance on the galactic XRBs) led to a set of weighting coefficients  {very similar to the four-class classifier}: 7.5 for location, 4.3 for spectrum, 2.6 for multiwavelength counterparts, and 6.1 for variability. {This may suggest that galactic XRBs are critical in the optimization of the four-class classifier on the reference sample, while most false positives in the test sample are extragalactic cases}. Once applied to the test sample, the classifier returns 79\% of AGN, 17\% of stars, 1.2\% of variable or young stars, 0.5\% of galactic XRBs, 1.6\% of extragalactic XRBs, and 0.6\% of CVs, which is again roughly consistent with the prior proportions.

\begin{table}[]
    \centering
    \caption{Confusion matrix and metrics of the classification with six classes, applied to the reference sample of 4XMM-DR10.}
    \begin{tabular}{c|cccccc}
        Truth $\rightarrow$ & AGN & Star & V/Y* & CV & gXB & eXB\\\hline
        $\rightarrow$ AGN & 18234 & 9 & 2 & 147 & 11 & 18\\
        $\rightarrow$ Star & 12 & 4605 & 137 & 5 & 2 & 2\\
        $\rightarrow$ V/Y* & 1 & 45 & 53 & 3 & 0 & 0\\
        $\rightarrow$ CV & 1 & 0 & 1 & 85 & 6 & 0\\
        $\rightarrow$ gXB & 2 & 0 & 1 & 2 & 55 & 0\\
        $\rightarrow$ eXB & 61 & 0 & 0 & 8 & 0 & 429\\
        Total & 18311 & 4659 & 194 & 250 & 75 & 449\\
        Total cl. & 18421 & 4659 & 101 & 60 & 487 & 99\\\hline 
        $recall$ & 99.6 & 98.8 & 27.3 & 34.0 & 73.3 & 95.5\\
        $precision$ & 97.6 & 93.6 & 68.6 & 87.9 & 95.6 & 95.3\\
        $f_1$ score & .986 & .961 & .391 & .490 & .830 & .954\\
    \end{tabular}
    \tablefoot{The first six rows show the number counts of objects classified as AGN, normal stars, variable or young stars, CVs, galactic XRBs, and extragalactic XRBs, respectively.  The last three rows give the $recall$ rate, the true positive rate ($precision$), and the $f_1$ score associated with each class.}
    \label{tab:9}
\end{table}

\subsubsection{Other model assumptions}

Our model assumes that the different classes given as parameters correspond to specific X-ray properties that enable them to be distinguished from one another. However, the $f_1$ score of XRBs is still lower than the one of AGN and stars, despite the maximization process: this may be due to the frequent overlap between AGN and XRB properties, whereas the AGN class is much more dominant in terms of number counts (and thus is favored by the priors). A subdivision of the XRB class into smaller, more homogeneous populations may partly solve this issue; however, these populations (e.g., black hole XRBs, separated further by spectral state) are not yet identified in sufficient number to constitute a reliable training sample. Future works of manual classification, such as the inspection of our XRB or CV candidates, will be able to validate some of them and allow a refinement of each class.

Another assumption of the model is that all multiwavelength counterparts are actual and contemporaneous associations, which is limited by our crossmatch algorithm. The spurious correlations problem {is fairly well addressed} by the use of the Bayesian cross-correlation algorithm NWAY, selecting the most likely counterpart for each source and assigning it a flag. To retrieve contemporaneous associations we would need a large overlap between X-ray and optical surveys, as will be the case of eROSITA and Vera C. Rubin surveys.

For the same purpose of improving the catalog enrichment, further improvements of our method could be to include constraining X-ray upper limits of each source. 
This could also enable the use of other variability properties, for example the maximum X-ray flux drop (or  increase) factor over 7 days (presumably short enough to not see important variations in AGN) or other time-dependent features. 

\section{Summary and outlook}

We developed a probabilistic classification scheme for X-ray sources based on their properties. It is one of the first classifiers to be applied to large fractions of current X-ray catalogs. We applied it to some of the latest versions of the \textit{Swift}-XRT and \textit{XMM-Newton} catalogs of unique sources, 2SXPS and 4XMM-DR10, after augmenting them with multi-instrument X-ray flux variability and multiwavelength data.

\begin{itemize}
\item[$\bullet$] After cross-correlating the X-ray catalog with catalogs of AGN, stars, XRBs, and CVs, we identified a reference sample of about 25000 known sources belonging to one of these classes.
\item[$\bullet$] We estimated the probability densities of each property for each class. Gathering properties in four categories, we computed for each source and each category the likelihood of each class. We applied weights to these likelihoods according to an optimization that maximizes the $f_1$ score of a chosen class. The Bayes formula was then applied using realistic priors to obtain the posterior probability of each class. As a result, the reasons for each classification are explicitly available, quantified by the influence of each property.
\item[$\bullet$] Evaluating the classifier on the reference sample, we obtain the best performance for the AGN and star classes, with over {97}\% of both $recall$ and $precision$. After optimizing the classifier to XRBs, we correctly retrieve more than 85\% of them, and we obtain less than 10\% of XRB false positives. Cataclysmic variables are less well classified because of their low prior and large diversity in properties.
\item[$\bullet$] We obtain consistent results when testing the classifier on the test samples, composed of more than 50\% of 2SXPS and 4XMM-DR10. By analyzing the classifier mistakes, the main limitations were identified, namely the lack of sufficient data, the loss of some multiwavelength counterparts, and discrepancies in the variability of some sources.
\item[$\bullet$] We implemented and tested an outlier measure to spot sources that belonged to none of the classes. Though applying a threshold on this measure prove to retrieve a substantial fraction of them, it selects in the majority of cases spurious or bad-quality sources, and some subclasses of each class with properties differing from the bulk of their class. Future studies will address a sample of remarkable outliers identified among unknown sources, as well as the classification of other X-ray catalogs, such as Chandra's CSC2 \citep{Evans2010}.
\end{itemize}

Further work could be done to improve the classifier performance and reduce its biases. This classifier can also be adapted to any other X-ray survey, it can be modified to include different source classes, and it can be enriched with other types of observables, according to the scientific interest. It will be useful in determining different homogeneous populations as well as identifying rare objects. 
We have released version 1.0 of the code and its user guide at the following address: \url{https://github.com/htranin/classificationXray}.


\begin{acknowledgements}
This research has made use of several tools and services that we acknowledge here: TOPCAT version 4.8 \citep{Taylor2005} ; the VizieR catalog access tool\footnote{https://vizier.u-strasbg.fr/viz-bin/VizieR}, CDS, Strasbourg, France \citep{vizier}; "Aladin sky atlas"\footnote{https://aladin.u-strasbg.fr/AladinLite/} developed at CDS, Strasbourg Observatory, France \citep{aladin2000, aladin2014}; the SIMBAD database\footnote{http://simbad.u-strasbg.fr/simbad/}, operated at CDS, Strasbourg, France \citep{simbad}; and Astropy\footnote{http://www.astropy.org}, a community-developed core Python package for Astronomy \citep{astropy2013, astropy2018}. We acknowledge support from the CNES and this project has received funding 
from the European Union's Horizon 
2020 research and innovation 
program under grant agreement 
n°101004168, the XMM2ATHENA project.
\end{acknowledgements}

\bibliographystyle{aa}
\bibliography{sample}

\begin{appendix}

\onecolumn

\section{Property distributions}

The properties used to infer our classifications, either provided in the X-ray catalogs or during catalog enrichment, offer different levels of separation between object types. We show in Fig. A.1 a more detailed version of Fig. 2, computed from the 4XMM-DR10 reference sample. Figure A.2 allows the distributions of the main properties to be compared, both between the reference and the test samples and between 2SXPS and 4XMM-DR10. They show a great similarity between the two catalogs, and the same biases when going from a reference sample to the corresponding test sample. The main biases are discussed in Sect. 4.2. 

\begin{figure}[h]
\centering
    \includegraphics[width=13cm]{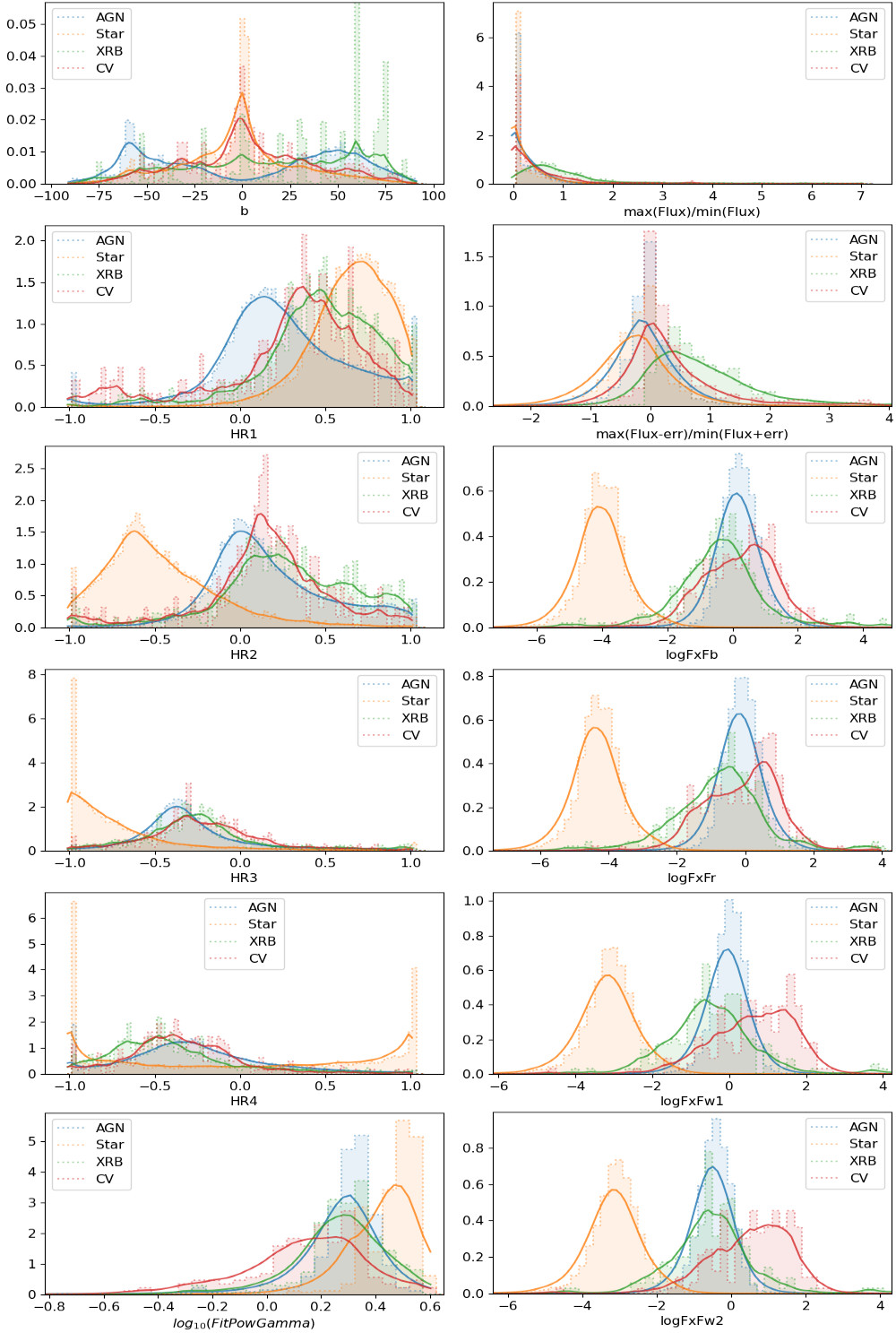}
    \caption{Distributions of all properties used in the classification, for the reference sample of 4XMM-DR10, and their KDE. From top to bottom, from left to right: galactic latitude ($b$), the four hardness ratios ($HR1, HR2, HR3, HR4$), index of the power-law fit, X-ray variability ratios, and X-ray to other wavelength flux ratios for the optical bands $b$ and $r$ and the infrared bands $W1$ and $W2$. The y axis is in arbitrary units.}
    \label{fig:A1}
\end{figure}

\begin{figure}
\centering
    \includegraphics[width=\textwidth]{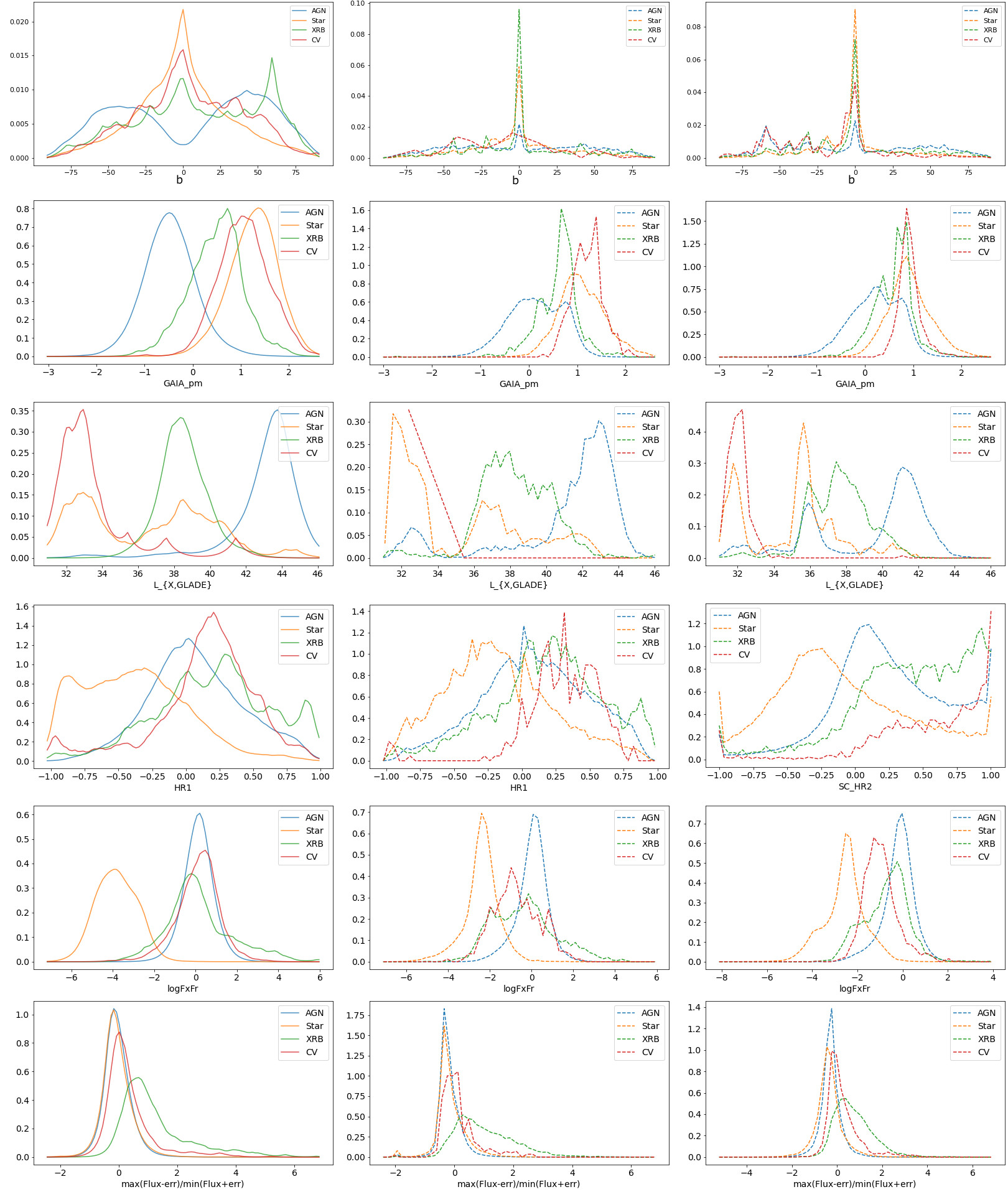}
    \caption{Distributions of different properties in the 2SXPS and 4XMM-D10 test samples (second and third columns; labels then stand for the selected class) and their KDE on the 2SXPS training sample (first column). From top to bottom: galactic latitude, Gaia proper motion, X-ray luminosity from GLADE, X-ray to r-band flux ratio, hardness ratio HR1, between \textit{Swift} soft (0.3--1 keV) and medium (1--2 keV) bands -- resp. HR2, between \textit{XMM-Newton} band 2 (0.5--1 keV) and band 3 (1--2 keV) -- and the lower limit on the variability ratio. The y axis is in arbitrary units.}
    \label{fig:my_master}
\end{figure}

%


\end{appendix}

\end{document}